\pdfoutput=1
\documentclass[11pt,a4paper]{article}
\def\isarxiv{1}
\input{diffPreamble}

\usepackage{ifthen} 
\newboolean{pdflatex}
\setboolean{pdflatex}{true} 

\newboolean{articletitles}
\setboolean{articletitles}{true} 

\newboolean{uprightparticles}
\setboolean{uprightparticles}{false} 

\newboolean{inbibliography}
\setboolean{inbibliography}{false} 

\usepackage[top=1in, bottom=1.25in, left=1in, right=1in]{geometry}

\usepackage{microtype}
\usepackage{lineno}  
\usepackage{xspace} 
\usepackage{caption} 

\usepackage{graphicx}  
\usepackage{color}
\usepackage{colortbl}
\graphicspath{{./figs/}} 

\usepackage{amsmath} 
\usepackage{amssymb}
\usepackage{amsfonts}
\usepackage{upgreek} 

\newcommand*\patchAmsMathEnvironmentForLineno[1]{%
\expandafter\let\csname old#1\expandafter\endcsname\csname #1\endcsname
\expandafter\let\csname oldend#1\expandafter\endcsname\csname
end#1\endcsname
 \renewenvironment{#1}%
   {\linenomath\csname old#1\endcsname}%
   {\csname oldend#1\endcsname\endlinenomath}%
}
\newcommand*\patchBothAmsMathEnvironmentsForLineno[1]{%
  \patchAmsMathEnvironmentForLineno{#1}%
  \patchAmsMathEnvironmentForLineno{#1*}%
}
\AtBeginDocument{%
\patchBothAmsMathEnvironmentsForLineno{equation}%
\patchBothAmsMathEnvironmentsForLineno{align}%
\patchBothAmsMathEnvironmentsForLineno{flalign}%
\patchBothAmsMathEnvironmentsForLineno{alignat}%
\patchBothAmsMathEnvironmentsForLineno{gather}%
\patchBothAmsMathEnvironmentsForLineno{multline}%
\patchBothAmsMathEnvironmentsForLineno{eqnarray}%
}

\usepackage{hyperxmp}

\usepackage[pdftex]{hyperref}

\usepackage[colorinlistoftodos,textsize=scriptsize]{todonotes}

\usepackage[bottom,flushmargin,hang,multiple]{footmisc}

\usepackage[all]{hypcap} 

\usepackage{xspace} 
\usepackage{upgreek}

\def\babar  {\mbox{BaBar}\xspace}
\def\belle  {\mbox{Belle}\xspace}

\def\MagUp {\mbox{\em Mag\kern -0.05em Up}\xspace}

\ifthenelse{\boolean{uprightparticles}}%
{
 
 \def\Pgamma      {\ensuremath{\upgamma}\xspace}

 \def\Pmu         {\ensuremath{\upmu}\xspace}                 
 \def\Pnu         {\ensuremath{\upnu}\xspace}                 
                  
 \def\Ppi         {\ensuremath{\uppi}\xspace}

 \def\Ptau        {\ensuremath{\uptau}\xspace}                 
                  
 \def\Pphi        {\ensuremath{\upphi}\xspace}

 \def\Ppsi        {\ensuremath{\uppsi}\xspace}

 \def\PDelta      {\ensuremath{\Delta}\xspace}                 
 \def\PXi      {\ensuremath{\Xi}\xspace}                 
 \def\PLambda      {\ensuremath{\Lambda}\xspace}                 
 \def\PSigma      {\ensuremath{\Sigma}\xspace}                 
 \def\POmega      {\ensuremath{\Omega}\xspace}                 
 \def\PUpsilon      {\ensuremath{\Upsilon}\xspace}                 
 
 \def\PB      {\ensuremath{\mathrm{B}}\xspace}                 
                  
 \def\PD      {\ensuremath{\mathrm{D}}\xspace}

 \def\PJ      {\ensuremath{\mathrm{J}}\xspace}                 
 \def\PK      {\ensuremath{\mathrm{K}}\xspace}

 \def\PW      {\ensuremath{\mathrm{W}}\xspace}

 \def\PZ      {\ensuremath{\mathrm{Z}}\xspace}                 
                  
 \def\Pb      {\ensuremath{\mathrm{b}}\xspace}                 
 \def\Pc      {\ensuremath{\mathrm{c}}\xspace}                 
 \def\Pd      {\ensuremath{\mathrm{d}}\xspace}                 
 \def\Pe      {\ensuremath{\mathrm{e}}\xspace}

 \def\Pi      {\ensuremath{\mathrm{i}}\xspace}

 \def\Pn      {\ensuremath{\mathrm{n}}\xspace}                 
                  
 \def\Pp      {\ensuremath{\mathrm{p}}\xspace}                 
 \def\Pq      {\ensuremath{\mathrm{q}}\xspace}                 
                  
 \def\Ps      {\ensuremath{\mathrm{s}}\xspace}                 
 \def\Pt      {\ensuremath{\mathrm{t}}\xspace}                 
 \def\Pu      {\ensuremath{\mathrm{u}}\xspace}

}
{
 
 \def\Pgamma      {\ensuremath{\gamma}\xspace}

 \def\Pmu         {\ensuremath{\mu}\xspace}                 
 \def\Pnu         {\ensuremath{\nu}\xspace}                 
                  
 \def\Ppi         {\ensuremath{\pi}\xspace}

 \def\Ptau        {\ensuremath{\tau}\xspace}                 
                  
 \def\Pphi        {\ensuremath{\phi}\xspace}

 \def\Ppsi        {\ensuremath{\psi}\xspace}                 
                  
 \mathchardef\PDelta="7101
 \mathchardef\PXi="7104
 \mathchardef\PLambda="7103
 \mathchardef\PSigma="7106
 \mathchardef\POmega="710A
 \mathchardef\PUpsilon="7107
                  
 \def\PB      {\ensuremath{B}\xspace}                 
                  
 \def\PD      {\ensuremath{D}\xspace}

 \def\PJ      {\ensuremath{J}\xspace}                 
 \def\PK      {\ensuremath{K}\xspace}

 \def\PW      {\ensuremath{W}\xspace}

 \def\PZ      {\ensuremath{Z}\xspace}                 
                  
 \def\Pb      {\ensuremath{b}\xspace}                 
 \def\Pc      {\ensuremath{c}\xspace}                 
 \def\Pd      {\ensuremath{d}\xspace}                 
 \def\Pe      {\ensuremath{e}\xspace}

 \def\Pi      {\ensuremath{i}\xspace}

 \def\Pn      {\ensuremath{n}\xspace}                 
                  
 \def\Pp      {\ensuremath{p}\xspace}                 
 \def\Pq      {\ensuremath{q}\xspace}                 
                  
 \def\Ps      {\ensuremath{s}\xspace}                 
 \def\Pt      {\ensuremath{t}\xspace}                 
 \def\Pu      {\ensuremath{u}\xspace}

}

\makeatletter
\ifcase \@ptsize \relax
  \newcommand{\miniscule}{\@setfontsize\miniscule{4}{5}}
\or
  \newcommand{\miniscule}{\@setfontsize\miniscule{5}{6}}
\or
  \newcommand{\miniscule}{\@setfontsize\miniscule{5}{6}}
\fi
\makeatother

\DeclareRobustCommand{\optbar}[1]{\shortstack{{\miniscule (\rule[.5ex]{1.25em}{.18mm})}
  \\ [-.7ex] $#1$}}

\def\electron   {{\ensuremath{\Pe}}\xspace}
\def\ep         {{\ensuremath{\Pe^+}}\xspace}
 
\def\epem       {{\ensuremath{\Pe^+\Pe^-}}\xspace}

\def\muon       {{\ensuremath{\Pmu}}\xspace}
\def\mup        {{\ensuremath{\Pmu^+}}\xspace}
\def\mun        {{\ensuremath{\Pmu^-}}\xspace} 
\def\mumu       {{\ensuremath{\Pmu^+\Pmu^-}}\xspace}

\def\taup       {{\ensuremath{\Ptau^+}}\xspace}
\def\taum       {{\ensuremath{\Ptau^-}}\xspace}

\def\ellm       {{\ensuremath{\ell^-}}\xspace}
\def\ellp       {{\ensuremath{\ell^+}}\xspace}
\def\ellell     {\ensuremath{\ell^+ \ell^-}\xspace}

\def\neu        {{\ensuremath{\Pnu}}\xspace}
\def\neub       {{\ensuremath{\overline{\Pnu}}}\xspace}

\def\g      {{\ensuremath{\Pgamma}}\xspace}

\def\W      {{\ensuremath{\PW}}\xspace}
\def\Wp     {{\ensuremath{\PW^+}}\xspace}
\def\Wm     {{\ensuremath{\PW^-}}\xspace}
\def\Wpm    {{\ensuremath{\PW^\pm}}\xspace}
\def\Z      {{\ensuremath{\PZ}}\xspace}

\def\quark     {{\ensuremath{\Pq}}\xspace}

\def\uquark    {{\ensuremath{\Pu}}\xspace}

\def\dquark    {{\ensuremath{\Pd}}\xspace}

\def\squark    {{\ensuremath{\Ps}}\xspace}

\def\cquark    {{\ensuremath{\Pc}}\xspace}
\def\cquarkbar {{\ensuremath{\overline \cquark}}\xspace}

\def\bquark    {{\ensuremath{\Pb}}\xspace}
\def\bquarkbar {{\ensuremath{\overline \bquark}}\xspace}

\def\tquark    {{\ensuremath{\Pt}}\xspace}

\def\pion   {{\ensuremath{\Ppi}}\xspace}
\def\piz    {{\ensuremath{\pion^0}}\xspace}

\def\pip    {{\ensuremath{\pion^+}}\xspace}
\def\pim    {{\ensuremath{\pion^-}}\xspace}

\def\kaon    {{\ensuremath{\PK}}\xspace}
  \def\Kbar    {{\kern 0.2em\overline{\kern -0.2em \PK}{}}\xspace}

\def\KorKbar    {\kern 0.18em\optbar{\kern -0.18em K}{}\xspace}
\def\Kz      {{\ensuremath{\kaon^0}}\xspace}

\def\Kp      {{\ensuremath{\kaon^+}}\xspace}
\def\Km      {{\ensuremath{\kaon^-}}\xspace}
\def\Kpm     {{\ensuremath{\kaon^\pm}}\xspace}

\def\KS      {{\ensuremath{\kaon^0_{\mathrm{ \scriptscriptstyle S}}}}\xspace}
\def\KL      {{\ensuremath{\kaon^0_{\mathrm{ \scriptscriptstyle L}}}}\xspace}
\def\Kstarz  {{\ensuremath{\kaon^{*0}}}\xspace}

\def\Kstar   {{\ensuremath{\kaon^*}}\xspace}

\def\Kstarp  {{\ensuremath{\kaon^{*+}}}\xspace}

  \def\Dbar    {{\kern 0.2em\overline{\kern -0.2em \PD}{}}\xspace}
\def\D       {{\ensuremath{\PD}}\xspace}

\def\DorDbar    {\kern 0.18em\optbar{\kern -0.18em D}{}\xspace}
\def\Dz      {{\ensuremath{\D^0}}\xspace}

\def\B       {{\ensuremath{\PB}}\xspace}
\def\Bbar    {{\ensuremath{\kern 0.18em\overline{\kern -0.18em \PB}{}}}\xspace}

\def\BorBbar    {\kern 0.18em\optbar{\kern -0.18em B}{}\xspace}
\def\Bz      {{\ensuremath{\B^0}}\xspace}
\def\Bzb     {{\ensuremath{\Bbar{}^0}}\xspace}
\def\Bu      {{\ensuremath{\B^+}}\xspace}
\def\Bub     {{\ensuremath{\B^-}}\xspace}
\def\Bp      {{\ensuremath{\Bu}}\xspace}
\def\Bm      {{\ensuremath{\Bub}}\xspace}

\def\Bd      {{\ensuremath{\B^0}}\xspace}
\def\Bs      {{\ensuremath{\B^0_\squark}}\xspace}
\def\Bsb     {{\ensuremath{\Bbar{}^0_\squark}}\xspace}
\def\Bdb     {{\ensuremath{\Bbar{}^0}}\xspace}
\def\Bc      {{\ensuremath{\B_\cquark^+}}\xspace}

\def\jpsi     {{\ensuremath{{\PJ\mskip -3mu/\mskip -2mu\Ppsi\mskip 2mu}}}\xspace}
\def\psitwos  {{\ensuremath{\Ppsi{(2S)}}}\xspace}

  \def\Y#1S{\ensuremath{\PUpsilon{(#1S)}}\xspace}

\def\proton      {{\ensuremath{\Pp}}\xspace}
\def\antiproton  {{\ensuremath{\overline \proton}}\xspace}
\def\neutron     {{\ensuremath{\Pn}}\xspace}

\def\Xires       {{\ensuremath{\PXi}}\xspace}

\def\Lz          {{\ensuremath{\PLambda}}\xspace}
\def\Lbar        {{\ensuremath{\kern 0.1em\overline{\kern -0.1em\PLambda}}}\xspace}
\def\LorLbar    {\kern 0.18em\optbar{\kern -0.18em \PLambda}{}\xspace}

\def\Lb      {{\ensuremath{\Lz^0_\bquark}}\xspace}
\def\Lbbar   {{\ensuremath{\Lbar{}^0_\bquark}}\xspace}

\def\Xib     {{\ensuremath{\Xires_\bquark}}\xspace}

\def\BF         {{\ensuremath{\mathcal{B}}}\xspace}

\def\BR         {\BF}
\newcommand{\decay}[2]{\ensuremath{#1\!\to #2}\xspace}         

\def\to                 {\ensuremath{\rightarrow}\xspace}

\def\qsq       {{\ensuremath{q^2}}\xspace}

\def\CP                {{\ensuremath{C\!P}}\xspace}

\def\Vtb  {{\ensuremath{V_{\tquark\bquark}}}\xspace}

\def\Vtss  {{\ensuremath{V_{\tquark\squark}^\ast}}\xspace}

\def\Vtbs  {{\ensuremath{V_{\tquark\bquark}^\ast}}\xspace}

\def\AFB      {\ensuremath{A_{\mathrm{FB}}}\xspace}

\def\AT#1     {\ensuremath{A_{\mathrm{T}}^{#1}}\xspace}           

\def\Bsmm     {\decay{\Bs}{\mup\mun}}
\def\Bdmm     {\decay{\Bd}{\mup\mun}}

\def\C#1      {\ensuremath{\mathcal{C}_{#1}}\xspace}                       
\def\Cp#1     {\ensuremath{\mathcal{C}_{#1}^{'}}\xspace}                    
\def\Ceff#1   {\ensuremath{\mathcal{C}_{#1}^{\mathrm{(eff)}}}\xspace}        
\def\Cpeff#1  {\ensuremath{\mathcal{C}_{#1}^{'\mathrm{(eff)}}}\xspace}       
\def\Ope#1    {\ensuremath{\mathcal{O}_{#1}}\xspace}                       
\def\Opep#1   {\ensuremath{\mathcal{O}_{#1}^{'}}\xspace}                    

\newcommand{\tev}{\ifthenelse{\boolean{inbibliography}}{\ensuremath{~T\kern -0.05em eV}\xspace}{\ensuremath{\mathrm{\,Te\kern -0.1em V}}}\xspace}
\newcommand{\gev}{\ensuremath{\mathrm{\,Ge\kern -0.1em V}}\xspace}
\newcommand{\mev}{\ensuremath{\mathrm{\,Me\kern -0.1em V}}\xspace}
\newcommand{\kev}{\ensuremath{\mathrm{\,ke\kern -0.1em V}}\xspace}
\newcommand{\ev}{\ensuremath{\mathrm{\,e\kern -0.1em V}}\xspace}
\newcommand{\gevc}{\ensuremath{{\mathrm{\,Ge\kern -0.1em V\!/}c}}\xspace}
\newcommand{\mevc}{\ensuremath{{\mathrm{\,Me\kern -0.1em V\!/}c}}\xspace}
\newcommand{\gevcc}{\ensuremath{{\mathrm{\,Ge\kern -0.1em V\!/}c^2}}\xspace}
\newcommand{\gevgevcccc}{\ensuremath{{\mathrm{\,Ge\kern -0.1em V^2\!/}c^4}}\xspace}
\newcommand{\mevcc}{\ensuremath{{\mathrm{\,Me\kern -0.1em V\!/}c^2}}\xspace}

\def\gsim{{~\raise.15em\hbox{$>$}\kern-.85em
          \lower.35em\hbox{$\sim$}~}\xspace}
\def\lsim{{~\raise.15em\hbox{$<$}\kern-.85em
          \lower.35em\hbox{$\sim$}~}\xspace}

\def\tell1  {TELL1\xspace}
\def\ukl1   {UKL1\xspace}

\usepackage{decays}
\usepackage{amsmath}
\newcommand{\aerr}[2]{{\:}^{+{\:}#1}_{-{\:}#2}}%

\usepackage{cite} 
\usepackage{mciteplus}

\usepackage{wrapfig}
\newcommand{\skipit}[1]{}

\usepackage{helvet}

\newcommand{\IF}[4]{\ifthenelse{\equal{#1}{#2}}{#3}{#4}}%
\newcommand{\arXiv}[2]{\IF{\isarxiv}{1}{#1}{#2}\xspace}
\newcommand{\mycite}[2][none]{\arXiv{~\cite{#2}}{\IF{#1}{none}{}{~\cite{#1}}}}

\usepackage{xfrac}

\usepackage{newfloat}
\usepackage{caption}
\DeclareFloatingEnvironment[fileext=frm,placement={!ht},name=Inset]{myfloat}

\usepackage{lipsum}
\usepackage[framemethod=TikZ]{mdframed}
\usepackage{multicol}

\newenvironment{frameenv}[2][tb]
    {\begin{myfloat}[#1]
    \begin{mdframed}[roundcorner=10pt,backgroundcolor=blue!10]
    \caption{\bfseries\boldmath #2}\centering \begin{minipage}{0.95\textwidth}\small
    } 
    {\end{minipage}\\\vskip 0.5em~\end{mdframed}\end{myfloat}
    }

\def\Wmst     {{\ensuremath{\PW^{\ast-}}}\xspace}

\begin{document}

\renewcommand{\thefootnote}{\fnsymbol{footnote}}
\setcounter{footnote}{1}


\begin{titlepage}

\vspace*{-1.5cm}

\noindent
\begin{tabular*}{\linewidth}{lc@{\extracolsep{\fill}}r@{\extracolsep{0pt}}}
\ifthenelse{\boolean{pdflatex}}
 & & Nikhef-2016-024 \\  
 & & \today \\ 
\hline
\end{tabular*}

\vspace*{4.0cm}

{\bf\boldmath\huge
\begin{center}
  Rare decays of $b$ hadrons
\end{center}
}

\vspace*{2.0cm}

\begin{center}
Patrick Koppenburg$^1$\footnote{Patrick.Koppenburg@cern.ch},{
Zden\v{e}k Dole\v{z}al$^2$, 
M\'{a}ria Smi\v{z}ansk\'{a}$^3$}
\bigskip\\
{\normalfont\itshape\footnotesize
$^1$Nikhef, Amsterdam, Netherlands\\
$^2$Charles University in Prague, Czech Republic\\
$^3$Lancaster University, United Kingdom\\
}
\end{center}

\vspace{\fill}

\arXiv{\begin{abstract}
  \noindent 
  Rare decays of \bquark hadrons provide a powerful way of identifying 
  contributions from physics beyond the Standard Model, in particular
  from new hypothetical particles too heavy to be produced at 
  colliders. The most relevant experimental measurements are reviewed and
  possible interpretations are briefly discussed.  
  \vskip 1em Contribution to Scholarpedia~\cite{Koppenburg:2016rji}.\footnote{\url{http://www.scholarpedia.org/article/Rare_decays_of_b_hadrons}.}
  This is the arXiv version with many more references. It corresponds to 
  the Summer 2020 revision.
\end{abstract}}{}

\vspace*{2.0cm}
\vspace{\fill}
{\footnotesize 
\centerline{\copyright Scholarpedia, licence \href{http://creativecommons.org/licenses/by-nc-sa/3.0/deed.en_US}{BY-NC-SA 3.0}.}}
\vspace*{2mm}
\end{titlepage}

\pagestyle{empty}  


\newpage
\setcounter{page}{2}
\mbox{~}


\renewcommand{\thefootnote}{\arabic{footnote}}
\setcounter{footnote}{0}

\tableofcontents
\section*{Document History}
\begin{description}
\item[June 2016:] Original version.
\item[September 2017:] Corrected mistake in \bgs branching fraction. Added \Bsmm results and prediction. New \decay{\Bs}{\taup\taum} search. New \blls results (both angular and $R_\Kstarz$) and theory predictions. 
New \decay{\bquark}{\cquark\taum\neub} results. Updated theory references for Wilson coefficient fits.
\item[October 2017:] Minor corrections including a corrected figure~\ref{Fig:BllKsAngles}.
\item[May 2019:] Added of new results on \Bmm and lepton universality by LHCb, ATLAS, CMS and Belle, and updated Fits section. Added a paragraph on \CP violation in \bgs decays.  Updated \blls results and figures.
\end{description}

\cleardoublepage


\pagestyle{plain} 
\setcounter{page}{1}
\pagenumbering{arabic}


%
\section{Introduction}\label{Sec:Intro}
Physics studies fundamental interactions and their effects. At the most basic
level, particle physics aims to describe the fundamental blocks of matter and
their interactions. 
A century of research has led to the Standard Model of Particle Physics. 
It relies on firm theoretical grounds unifying quantum mechanics, special relativity and 
field theory, and is successful at describing all phenomena measured in particle 
interactions, whether at low or very high energies.
Yet, it has an empirical character with many parameters that need to be determined by experiment, and 
it is incomplete as it does not account for gravity, does not explain the baryon 
asymmetry in the Universe
and does not provide a candidate for dark matter. The Standard Model is therefore believed 
to be an approximation of a more complete theory that is currently unknown (just like Newton's 
laws are an approximation of General Relativity). The primary goal of research in particle 
physics is to find this more complete theory. In the following ``New Physics'' is used 
as a catch-all for 
any contribution --- usually associated with a new particle --- not included in the Standard Model.

Rare decays of hadrons containing  a heavy ``beauty'' 
(also called ``bottom'') quark, 
denoted \bquark hadrons, provide a powerful way of exploring yet 
unknown physics. Small contributions from virtual new particles that are too heavy to be 
produced at colliders may lead to measurable deviations from the expected properties in the 
Standard Model.
See Inset~\ref{Ins:Virtual} for an example of a virtual particle.
\begin{frameenv}[b]{Virtual Particles}\label{Ins:Virtual}
\begin{wrapfigure}{l}{0.42\textwidth}
\begin{minipage}{0.4\textwidth}
\includegraphics[width=\textwidth]{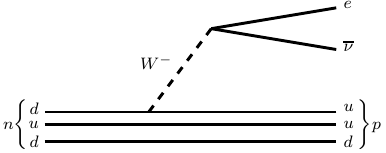}\hskip 0.09\textwidth
\caption{Feynman diagram of the beta decay of the neutron.}\label{Fig:beta}
\vskip -2em ~
  \end{minipage}
\end{wrapfigure}
Fundamental particle interactions are mediated by force carriers: The photon for the
electromagnetic interaction, the gluon for the strong interaction and the \Wpm and \Z bosons
for the weak interaction. In the nuclear beta decay a neutron decays to a proton,
an electron and a neutrino (Fig.~\ref{Fig:beta}). 
This weak interaction is mediated by a \emph{virtual} \Wm boson, sometimes denoted \Wmst.
Here \emph{virtual} means that the process violates
energy and momentum conservation for a very short time, as allowed by Heisenberg's 
Uncertainty Principle. The mass of the \Wm boson is about $80\:\gevcc$, while the 
neutron-proton mass difference is considerably less: $1.3\:\mevcc$. 
It is said that the \Wm boson is  ``off-shell''.
\end{frameenv}


The study of rare decays is an active field within flavour physics, the field of research 
studying transitions of quarks or leptons from one species (or ``flavour'') to another.
This article focuses on rare decays of hadrons containing \bquark 
quarks. The most prevalently produced  \bquark hadrons are the \Bz meson
composed of a \bquarkbar anti-quark and a \dquark quark, 
the \Bp (\bquarkbar{}\uquark) and \Bs (\bquarkbar{}\squark) mesons,
as well as the \Lb (\bquark{}\uquark{}\dquark) baryon. Their masses
are in the range $5$ to $6\:\gevcc$, which is about six times that of the proton,
but well below the mass of the \W boson of $80\:\gevcc$.
The corresponding
antiparticles \Bzb, \Bm, \Bsb and \Lbbar are obtained by replacing all
quarks by anti-quarks and vice-versa. 
The study of \CP violation involves investigation of differences in the behaviour of 
particles and antiparticles, and is the subject of a dedicated review in Ref.~\cite{Buras:2015}.  
The inclusion of charge conjugate processes is implied throughout this document.

Hadrons with \bquark quarks decay most of the 
time via a \decay{\bquark}{\cquark\Wmst} transition, where the asterisk indicates the 
\W boson is virtual.
The transitions \decay{\bquark}{\uquark\Wmst} also occur, but are less likely. 
These two transitions are called ``tree decays'' as the process involves a 
single mediator, the \Wm boson. An example of a tree \decay{\dquark}{\uquark\Wmst} transition
is shown in Fig.~\ref{Fig:beta}.

This article describes transitions involving more complicated processes. 
The quark transitions \decay{\bquark}{\dquark} and \decay{\bquark}{\squark}
do not happen at tree level in the Standard Model as the \Z boson
does not couple to quarks of different flavour.

\begin{wrapfigure}{r}{0.37\textwidth}
\vskip -1em 
\hskip 0.02\textwidth
\begin{minipage}{0.35\textwidth}
\includegraphics[width=\textwidth]{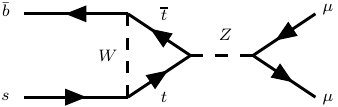}
\caption{Feynman diagram of the dominating Standard Model contribution 
to the decay \Bsmm.}\label{Fig:Bsmm1}
  \end{minipage}
\vskip -1em 
\end{wrapfigure}
Processes like the rare decay \Bsmm proceed via 
loops, as shown in Fig.~\ref{Fig:Bsmm1} (sometimes referred to as penguins, 
a word coined by John Ellis\mycite[symmetry-penguins]{Ellis:1977uk,symmetry-penguins}). 
Such processes are rare as the probability of a transition rapidly decreases 
with the number of electroweak vertices: two in the case of a tree decay, 
three or four for a loop.
Also, the heavier the virtual particles involved, the more suppressed the decay.
In the following, decays with probabilities in the range $10^{-4}$ to $10^{-10}$ are 
discussed.

Some of the most interesting decays are described in Section~\ref{Sec:Players}. They all 
have in common the following features:
\begin{enumerate}\setlength\itemsep{-0em}%
\item Suppressed decay amplitudes, as predicted by the Standard Model, which may potentially be of the same
  size as New Physics amplitudes.
\item Sufficiently precise Standard Model predictions for their decay rate, or any other observable of interest.
\item Experimental precision which potentially allows disentangling the Standard Model 
  contribution from other contributions. 
\end{enumerate}
A historical example, the decay \decay{\KL}{\mumu}, is described in Inset~\ref{Ins:KLmm}.

\begin{frameenv}[b]{\boldmath A Historical Example: \decay{\KL}{\mumu}}\label{Ins:KLmm}
\begin{wrapfigure}{r}{0.55\textwidth}
\vskip -0.5em 
\hskip 0.02\textwidth
\begin{minipage}{0.53\textwidth}
\includegraphics[width=0.45\textwidth]{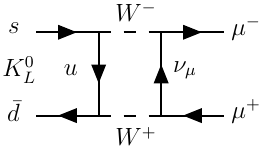}\hskip 0.09\textwidth
\includegraphics[width=0.45\textwidth]{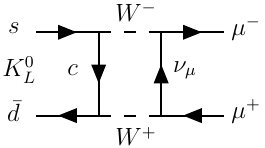}
\caption{Feynman diagrams of the two loops contributing to the decay \decay{\KL}{\mumu}.}\label{Fig:KLmm}
  \end{minipage}
\vskip -1.5em ~
\end{wrapfigure}
The \decay{\KL}{\mumu} decay is forbidden at tree level and 
had an important role in opening the field of rare decays
in the 1960s. Its unexpected non-observation 
allowed the prediction of the then unknown charm quark by 
Glashow, Iliopoulos and Maiani (``GIM mechanism'') in 
1970\mycite[Iliopoulos:2010]{Glashow:1970gm}. 
The idea of the GIM mechanism is that this decay only occurs via loops, one
involving the \uquark quark and the other the \cquark quark (Fig.~\ref{Fig:KLmm}). The 
amplitudes of the two loops are of opposite sign, causing complete
cancellation in the limit of equal up and charm quark masses.
The non-observation of this
decay could be explained by adding a new particle to the
theory, the \cquark quark, which was eventually discovered in 
1974\mycite{Aubert:1974js,Augustin:1974xw}.
This is an example of an observation of New Physics mediated by a new virtual 
particle. 
The \decay{\KL}{\mumu}  branching fraction is now measured
to be $(6.84\pm0.11)\times10^{-9}$\mycite{PDG}.
Nowadays there is a great deal of interest in the \Bs-counterpart of this decay: 
\Bsmm, discussed in Section~\ref{Sec:Bsmm}. 
\end{frameenv}


A special category of rare decays is those forbidden in the Standard Model, like 
lepton- or baryon-number violating decays. In their case the Standard Model 
prediction is effectively zero, 
but other models may predict non-zero rates. Any observation would be a sign of
New Physics.

Given the impressive success of the Standard Model, New Physics amplitudes  are 
known to be small. Therefore, any search for potentially observable deviations 
from Standard Model predictions will be facilitated if the Standard Model 
amplitudes are also suppressed, which is the case in rare decays. 
Studies of such decay modes require require large data 
samples to produce enough of the relevant particles. 
This is referred to as the {\it intensity frontier}, as opposed to the {\it energy frontier}
aiming at producing and studying heavy particles on-shell. 
The main experiments are briefly described in Section~\ref{Sec:Experiments}.

\subsection{Key players}\label{Sec:Players}
The description of the process of the formation of hadrons out of quarks and gluons, called 
hadronisation, is difficult and leads to large theoretical uncertainties. Theoretically favoured
are thus decays to purely leptonic final states, such as the decay \Bsmm (Section~\ref{Sec:Bsmm}). 
There is also interest in the charged counterparts of 
these decays, notably \decay{\Bp}{\ellp\neu}, where \ellp is any lepton, \ep, \mup, \taup
(Section~\ref{Sec:Blnu}). They are generated 
by a charged \Wp current, but have interesting theoretical connections to decays that are 
induced by loops.

Radiative \decay{\bquark}{\squark\gamma}, and  
electroweak penguin
\decay{\bquark}{\squark\ellell} and
\decay{\bquark}{\squark\neu\neub} decays are also of great interest.
These are quark-level transitions,
which cannot be measured directly as the quarks form immediately hadrons.
In experiments exclusive decays are detected,
and the inclusive decay is the sum of all contributions. For instance the 
decay \bgs was first observed by its exclusive contribution \BgKs (Section~\ref{Sec:bgs}).

Exclusive decays are experimentally favoured, but come with larger theoretical
uncertainties. The decay \BdllKs is a well-known example. While the decay rate is hard 
to compute
precisely, observables describing angular distributions of the decay products 
can be more precisely predicted (Section~\ref{Sec:BllKs}).

Among forbidden decays, lepton flavour violating decays of \bquark and \cquark hadrons,
like \decay{\B^0_{(s)}}{e^\pm\mu^\mp} or \decay{\Bp}{\Kp e^\pm\mu^\mp}, or of leptons,
like \decay{\mup}{\ep\gamma}, \decay{\tau^+}{\mu^+\mumu} or \decay{\tau^+}{\mu^+\gamma} 
are actively being searched for.
Rare charm hadron decays are also being studied, but the experimental 
sensitivity is presently not sufficient to reach the very low 
rates predicted in the Standard Model.
Finally, research in rare kaon decays is ongoing, 
though mainly at different experiments than those studying
rare charm or beauty hadron decays\mycite{PDG}. 
These channels are not further discussed in this article.
\arXiv{Other recent reviews on rare decays can be found in
  Refs.~\cite{Blake:2015tda,Borissov:2013yha}.}{}

\section{Theory}\label{Sec:Hamilton}\label{Sec:Theory}
This section describes briefly the theoretical framework that is commonly used 
to study rare decays. 
Its main goal is to define some vocabulary which is
commonly used in publications on rare decays.
It may be skipped by readers mostly interested in experimental results. 

The common theoretical approach to rare decays is model independent. 
In flavour physics and in particular in rare decays studies,
the underlying physics is parametrised in terms of an effective Hamiltonian
describing the transition amplitude of an initial state $I$ 
to a final state $F$ following Fermi's Golden Rule~\cite{Dirac243,Fermi:1949}. The partial decay width is written as
\begin{equation}\label{eq:Fermi}
  \Gamma(\decay{I}{F}) = \frac{2\pi}{\hbar}\left|\left<F|{\cal H}_\text{eff}|I\right>\right|^2\times\text{phase-space}.
\end{equation}
Experimentally, the branching fraction $\BR$ is measured rather than the decay width. They are related by
\begin{equation}\label{eq:B2I}
  {\BR(\decay{I}{F}) = \frac{\Gamma(\decay{I}{F})}{\Gamma(I,\text{total})}},\quad \Gamma(I,\text{total}) = \frac{1}{\tau_I},
\end{equation}
where $\tau_I$ is the lifetime of particle $I$ 
(and natural units with $c=\hbar=1$ are used).

The Standard Model prediction for any particular transition can be inferred from a 
calculation of the effective Hamiltonian derived from the Standard Model Lagrangian. 
This Hamiltonian is parametrised in terms of a sum of operators 
${\cal O}_i$ and Wilson coefficients $C_i$
\begin{equation}\label{eq:OPE}
  {\cal H}_\text{eff} = -\frac{G_F}{\sqrt{2}}\sum\limits_i V_\text{CKM}C_i{\cal O}_i,
\end{equation}
where $V_\text{CKM}$ stands for some product of Cabibbo-Kobayashi-Maskawa 
matrix elements that describe the probability of 
given transitions between different quark flavours. 
The operators encompass the information about the
Lorentz structure 
and the Wilson coefficients encode the effects of higher energy scales. 
In the case of the Standard Model these are the effects of the \W, \Z bosons and 
top quarks, which are effectively removed from the theory and incorporated in
the coefficients.

Any \decay{I}{F} decay can be described by this effective Hamiltonian,
usually with many terms begin irrelevant, with $\left<F|{\cal O}_i|I\right>=0$. Thus studying a set of decays will give 
various constraints on the effective Hamiltonian, permitting global fits
to Wilson coefficients. This is briefly discussed in Section~\ref{Sec:Fits}.
This procedure does not simplify the computation of the amplitudes,
as the matrix elements $\left<F|{\cal O}_i|I\right>$ contain the most difficult parts
of the calculation. It provides however a common language that is not dependent
on the considered New Physics model.

In particular, calculations of decay rates of exclusive decays with hadrons
in the final state (\BdmmKs for example) are difficult.
Our lack of knowledge needs to be parametrised in heuristic
quantities that describe the hadronisation,
like form-factors and decay constants. They can be calculated in lattice QCD 
and, in many cases, can also be determined experimentally. Their discussion is beyond the 
scope of this document.

\begin{wrapfigure}{r}{0.33\textwidth}
\begin{minipage}{0.3\textwidth}
\includegraphics[width=\textwidth]{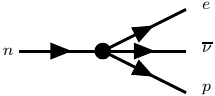}\hskip 0.09\textwidth
\caption{Effective process \decay{\neutron}{\proton\electron\neub} for the 
nuclear beta decay.}\label{Fig:Fermi}
  \end{minipage}
\end{wrapfigure}
The operators ${\cal O}_{1,2}$ describe the $V-A$ structure of weak decays and 
first-order corrections. For example, the \W boson having been absorbed into the $C_1$ and $C_2$ 
coefficients, the nuclear beta decay \decay{\neutron}{\proton\electron\neub} 
is represented by a four-fermion operator as shown in Fig.~\ref{Fig:Fermi}. This
is how Enrico Fermi first described the process in 1934~\cite{Fermi:1934hr}. 
The operators ${\cal O}_{3\text{--}6,8}$
describe loops involving gluons. They are 
not of interest for this article.

Of most interest 
in rare decays are the suppressed operators ${\cal O}_{7}$, ${\cal O}_{9}$ and 
${\cal O}_{10}$. 
The operator ${\cal O}_{7}$  dominates the radiative decay
\decay{\bquark}{\squark\gamma} giving a decay width
\begin{equation}\label{eq:bgs}
  \Gamma(\bgs) = \frac{G_F^2\alpha_{\rm EM}m_\Pb^5}{32\pi^4}|\Vtss\Vtb|^2 |C_{7}|^2 
  + \text{corrections},
\end{equation}
where $\alpha_{\rm EM}$ is the electromagnetic constant, $m_b$ the \bquark quark mass,
and $V_{ij}$ are parameters of the CKM matrix. A measurement of the \bgs branching fraction
thus provides a direct constraint on  $C_7$.

The operators ${\cal O}_{9}$ and ${\cal O}_{10}$ dominate \decay{\bquark}{\quark\ell\ell} transitions,
with ${\cal O}_{9}$ corresponding to a vector and ${\cal O}_{10}$ to an axial current.
Finally the decays \decay{\B}{\ellell} are, in the Standard Model, dominated by 
operator ${\cal O}_{10}$, with 
a decay rate which can be written as\mycite{Bobeth:2013uxa}
\begin{equation}\label{eq:Bll}
  {\Gamma}(\decay{B}{\ellell}) = \frac{G_F^2M_W^2m_B^3f_B^2}{8\pi^5}|\Vtbs V_{tq}|^2\frac{4m_\ell^2}{m_B^2}\sqrt{1-\frac{4m_\ell^2}{m_B^2}} |C_{10}|^2 + \text{corrections},
\end{equation}
for $B=\Bd,\Bs$ (and $q=d,s$) with
$f_B$ the \B decay constant and
$V_{ij}$ CKM matrix elements. 
\arXiv{It is to be noted that in the \Bs case 
only the heavy mass eigenstate contributes, and hence the $B_s^{H}$ decay width
must be used to compute the branching fraction\mycite{DeBruyn:2012wj,DeBruyn:2012wk}.}{} 
The \Bsmm branching fraction thus provides a constraint on $C_{10}$. 
Other operators, labelled ${\cal O}_\text{P}$ and ${\cal O}_\text{S}$, 
which are negligible in the Standard Model, could also
contribute to this decay.

If the $V-A$ structure of weak interactions is not assumed, new primed operators with 
flipped helicities appear, most notably ${\cal O}_7'$, and its Wilson coefficient
$C_7'$ which generate a right-handed photon in \bgs decays.

For a comprehensive review of the effective Hamiltonian used to study rare decays, 
see Refs.~\cite{Buchalla:1995vs,Buras:1997fb}. A more 
pedagogical introduction can be found in Chapter 20 of Ref.~\cite{Branco:1999fs}. 
Standard Model expectations of Wilson coefficients and operators 
have been calculated at next-to-leading order or 
better\mycite{Bobeth:1999mk,Bobeth:2003at,Huber:2005ig}.

There exist many theories beyond the Standard Model providing predictions for Wilson coefficients.
Often these values depend on unknown parameters of the theory, as masses of yet unseen new particles.
This is particularly the case for supersymmetry, 
a well-motivated extension of the Standard Model.

\section{Experiments}\label{Sec:Experiments}
There are essentially two families of experiments studying \bquark hadrons: 
\begin{description}
\item[\boldmath\B factories] are experiments based at \epem colliders operating most of the time at
  a collision energy near $10.6\gev$, corresponding to the mass of the $\PUpsilon(4S)$ resonance, 
  the lightest meson decaying to two \B mesons. 
  ARGUS\mycite{Albrecht:1988vy} at DESY (Hamburg, Germany),
  CLEO\mycite{Andrews:1982dp} at Cornell (Ithaka, USA),
  \babar\mycite{Aubert:2001tu} at SLAC (Stanford, USA),
  Belle\mycite{Abashian:2000cg} and its successor Belle II\mycite{Abe:2010gxa} at KEK (Tsukuba, Japan) 
  are notable examples of such experiments.
\item[Hadron collider experiments] operate at a \proton{}\proton or \proton{}\antiproton
  collider with centre-of-mass energies of several \tev.
  CDF and D0 were located at Fermilab's Tevatron (Batavia, USA)\mycite{Paulini:1999px}.
  ATLAS\mycite[Dunford:2014]{Dunford:2014,Aad:2008zzm}, 
  CMS\mycite{Chatrchyan:2008aa}
  and LHCb\mycite{Alves:2008zz}
  presently operate at CERN's LHC (Geneva, Switzerland).  
\end{description}
Hadron colliders have the advantage of much larger production rates: 
the production cross-section of \bquark quarks is 
a factor 500\:000 larger\mycite{LHCb-PAPER-2015-037} at the LHC than at a \B factory.
The advantage of the \B factories is cleanliness. 
Collision events with a produced $\PUpsilon(4S)$ resonance
are easy to identify, allowing for high efficiencies and low background levels. In such events
only two \B mesons are produced, making the reconstruction of the full collision 
event possible. In a typical LHC collision only one in hundred collisions produce a 
\bquark quark pair and the two \bquark hadrons are surrounded by hundreds 
of other particles. Efficient background fighting techniques are thus essential, but
have a cost in terms of efficiencies. Excellent background rejection
is achieved by precise vertexing and exploiting the 
\bquark-hadron flight distance.

The physics programme is also somewhat different:
\B factories have only access to \Bz and \Bp mesons (and \Bs mesons when operating at the 
$\PUpsilon(5S)$ resonance), while hadronic collisions produce all \bquark hadrons, including
the \Bs meson, the \Bc meson (composed of a \cquark quark and an \bquarkbar anti-quark),
as well as the \Lb and \Xib baryons.

\subsection[Short history of \bquark-quark physics]{\boldmath Short history of \bquark-quark physics}
\begin{figure}[t]
\centering
\includegraphics[width=\textwidth]{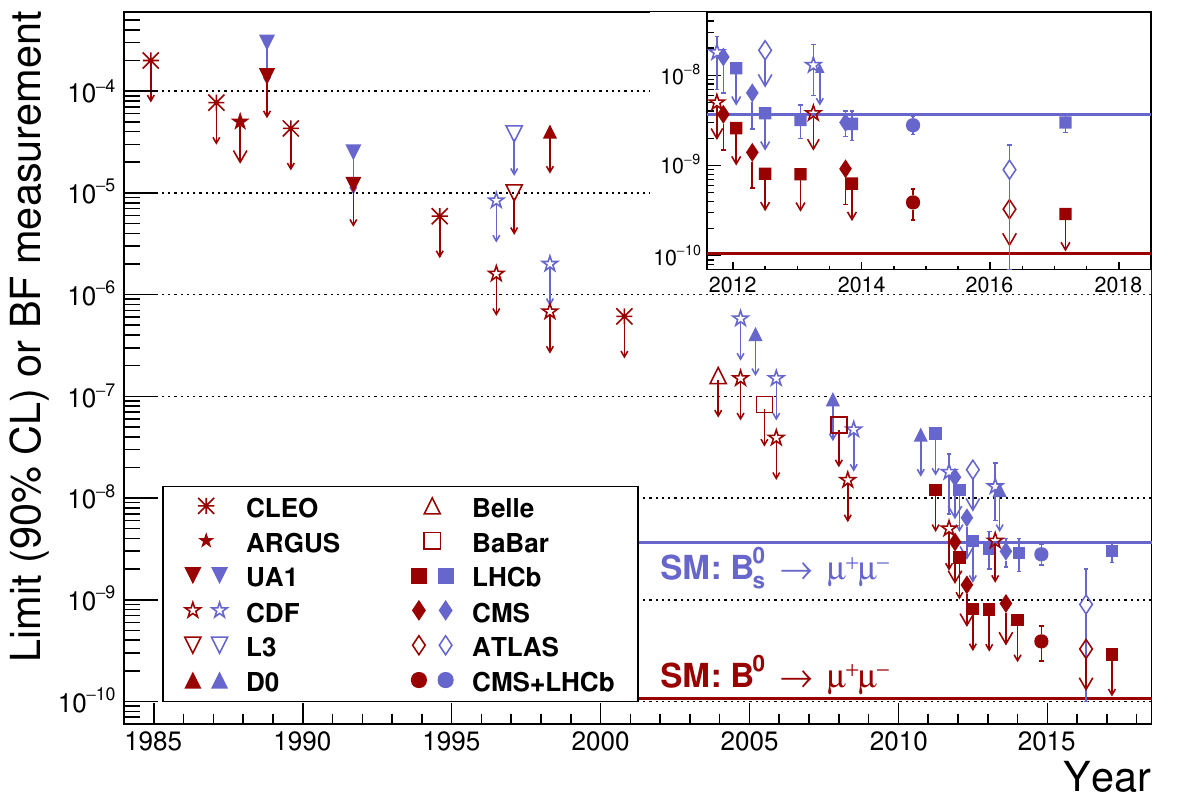}
\caption{History of \Bsmm and \Bdmm limits up to the \Bsmm observation. 
Figure courtesy of F. Dettori.
}
\label{Fig:BsmmHistory}
\end{figure}

After the discovery of the \bquark quark at Fermilab through the observation 
of mesons formed by a \bquark and an \bquarkbar anti-quark in 
1977\mycite{Herb:1977ek} and of 
the \B meson at Cornell\mycite{Bebek:1980zd,Behrends:1983er}, 
searches for 
rare decays of \bquark hadrons rapidly took pace. The first limit on
the decay \Bdmm was set by the CLEO collaboration in 1985\mycite{Giles:1984yg}, 
the start of a long quest during which the sensitivity was improved by 
six orders of magnitude, as illustrated in Fig.~\ref{Fig:BsmmHistory}.

The CLEO and ARGUS experiments
were located at \epem colliders operating at the $\PUpsilon(4S)$ resonance. 
The same concept was employed
and improved by the \babar and 
Belle experiments in the first decade of the 21st century.
If Cornell was initially able to produce few tens of 
\B\Bbar pairs per day, the PEP-II and KEKB accelerators at SLAC and KEK 
achieved a daily rate of one million \B\Bbar pairs.
In the meantime, experiments at CERN's LEP \epem
collider\mycite{Decamp:1990jra,Aarnio:1990vx,L3:1989aa,Ahmet:1990eg} and at 
Fermilab's Tevatron proton-anti-proton collider used higher energy collisions
to produce and study all \bquark-hadron species\mycite{Barker:2010iva,Rowson:2001cd}.
All the above-mentioned experiments have terminated their programme but most 
still exploit their data set to produce new results.
Belle and the associated accelerator complex 
is has been undergoing a major upgrade and has recently come back as 
the Belle II experiment\mycite{BrowderLP}.

\subsection{Present}
Until Belle II reaches its design luminosity,
the leadership in \bquark physics is taken by the LHCb 
experiment. Important contributions also come from 
the ATLAS and CMS experiments. These three experiments exploit LHC data
collected in proton-proton collisions at centre-of-mass energies of
$7$ (2010--11), $8$ (2012) and $13$\tev (2015--18).

ATLAS and CMS are detectors optimised for 
high-energy processes, such as the discovery of the Higgs 
boson\mycite[Jakobs:2015]{Aad:2012tfa,Chatrchyan:2012ufa}. They also perform 
\bquark-physics research,
most effectively in decays of \bquark hadrons to pairs of muons. This distinct signature allows
for efficient selection of these decays during the online filtering phase where a 
large reduction of the recorded collision rate is required, which is difficult to 
achieve for decays to electrons or hadrons.

The LHCb experiment on the contrary is optimised for the 
physics of hadrons containing \bquark and \cquark quarks.
It is a single-arm forward detector designed to exploit the relatively large $b\bar{b}$ 
production in LHC proton-proton collisions
in the forward direction. It includes a tracking system
surrounding a dipole magnet whose polarity can be reversed, silicon sensors coming 
as close as 8\:mm to the proton beam and a particle identification system based on 
Cherenkov radiation. The high-resolution silicon system exploits the typical \bquark-hadron 
flight distances of a few millimetres before their decay to select them. This sets
requirements on the number of $pp$ collisions per bunch crossing, defining an upper limit
to the total collision rate at which the experiment can operate. 
Consequently, the luminosity is decreased compared
to ATLAS and CMS.

\section{Main experimental results}\label{Sec:Results}
This section presents the main recent experimental results and their
interpretation. It starts with a more historical 
section on the decay \bgs which had (and still has) an important role
in the development of the field.
\subsection[The decay \bgs]{\boldmath The decay \bgs}\label{Sec:bgs}
\begin{wrapfigure}{r}{0.37\textwidth}
\vskip -5em 
\hskip 0.02\textwidth
\begin{minipage}{0.35\textwidth}
\includegraphics[width=\textwidth]{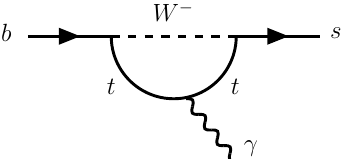}
\caption{Feynman diagram of the dominating Standard Model contribution 
to the decay \bgs.}\label{Fig:bgs}
  \end{minipage}
\end{wrapfigure}
In the Standard Model the decay \bgs occurs dominantly via a loop involving 
the top quark and the \W boson (Fig.~\ref{Fig:bgs}). 
It has played a very important role in flavour physics from the 
1980s\mycite{Campbell:1981rg}. At the time it was the dependency of the 
branching fraction on the then unknown top quark mass that was the driving 
force behind the 
theoretical calculations and the experimental searches
(the top quark mass affects the value of the $C_7$ coefficient in Eq.~\ref{eq:bgs}).
When \Bd--\Bdb mixing was (at the time surprisingly) observed in 1987, 
it became clear that the top quark was very heavy.
The top quark was eventually discovered at the Tevatron\mycite{Abe:1995hr} in 1995 and 
its mass measured, 
which determined the 
Standard Model decay rate of \bgs to be a few $10^{-4}$. 

\begin{wrapfigure}{l}{0.47\textwidth}
\includegraphics[width=0.45\textwidth]{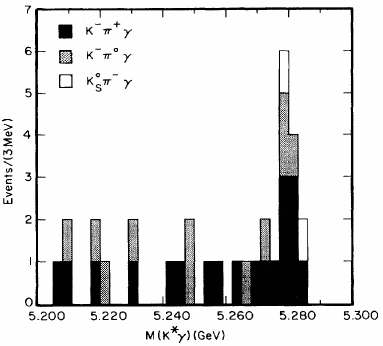}\hskip 0.05\textwidth%
\caption{First observation of the decay \decay{\Bd}{\Kstarz\gamma} by the CLEO %
collaboration (see the peak at 5.28\:\gev)~\cite{Ammar:1993sh}.}%
\label{Fig:B2KstarGamma}%
\vskip -1.5em ~
\end{wrapfigure}%
The first observation of the \bgs decay actually preceded the top quark observation. In 1993
the CLEO collaboration reported a signal of the exclusive decay \decay{\B}{\Kstar\gamma} 
with a branching fraction of $(4.5 \pm 1.5 \pm 0.9) \times 10^{-5}$\mycite{Ammar:1993sh}
(Fig.~\ref{Fig:B2KstarGamma}), where the first uncertainty is statistical and the second systematic.
This opened the quest for the inclusive decay \bgs,
i.e. the sum of all exclusive contributions. The branching fraction of 
this decay is more precisely calculable than its individual exclusive 
components, like \decay{\Bd}{\Kstarz\gamma}, allowing for more precise
comparisons of experimental and theoretical results. The experimental
challenges are discussed in more detail in Inset~\ref{Ins:bgs}

In the Standard Model, the left-handed chirality structure of the weak interactions makes the 
photon emitted in \bgs decays mainly left-handed. It is interesting to also probe right-handed 
contributions (sensitive to the Wilson coefficient $C_7'$), 
which requires determination of the polarisation of the photon. This is
challenging as the helicity (or chirality) of the photon cannot be measured directly in
the detector. Several methods have been proposed, none of which provides 
a strong constraint so far.  The first and so far only measurement of a non-zero
photon polarisation uses the decay
\decay{\B}{\Kp\pim\pip\Pgamma}\mycite{LHCb-PAPER-2014-001}, 
but the interpretation
in terms of the photon chirality is still unclear.
The most stringent constraints come from global fits to Wilson coefficients
(see Section~\ref{Sec:Fits}).

\begin{frameenv}[t]{\boldmath The inclusive \bgs spectrum}\label{Ins:bgs}
Many experiments located at \epem colliders 
have performed measurements using different methods. The total rate of any \B meson
to a photon plus anything (where the photon is not caused by an electromagnetic decay, 
e.g. \decay{\piz}{\gamma\gamma} or \decay{\eta}{\gamma\gamma})
can be measured by a sum of exclusive decay modes\mycite{Adriani:1993yn,Abe:2001hk,Aubert:2005cua,Aubert:2006gg,Aubert:2008av,delAmoSanchez:2010ae,Lees:2012wg,Saito:2014das}. 
A fully inclusive approach is also possible but more challenging\mycite{Alam:1994aw,Adam:1996ts,Barate:1998vz,Chen:2001fja,Koppenburg:2004fz,Lees:2012ym,Aubert:2007my,Limosani:2009qg}.

\begin{wrapfigure}{r}{0.47\textwidth}
\vskip -1.5em 
\hskip 0.02\textwidth
\begin{minipage}{0.45\textwidth}
\includegraphics[width=\textwidth]{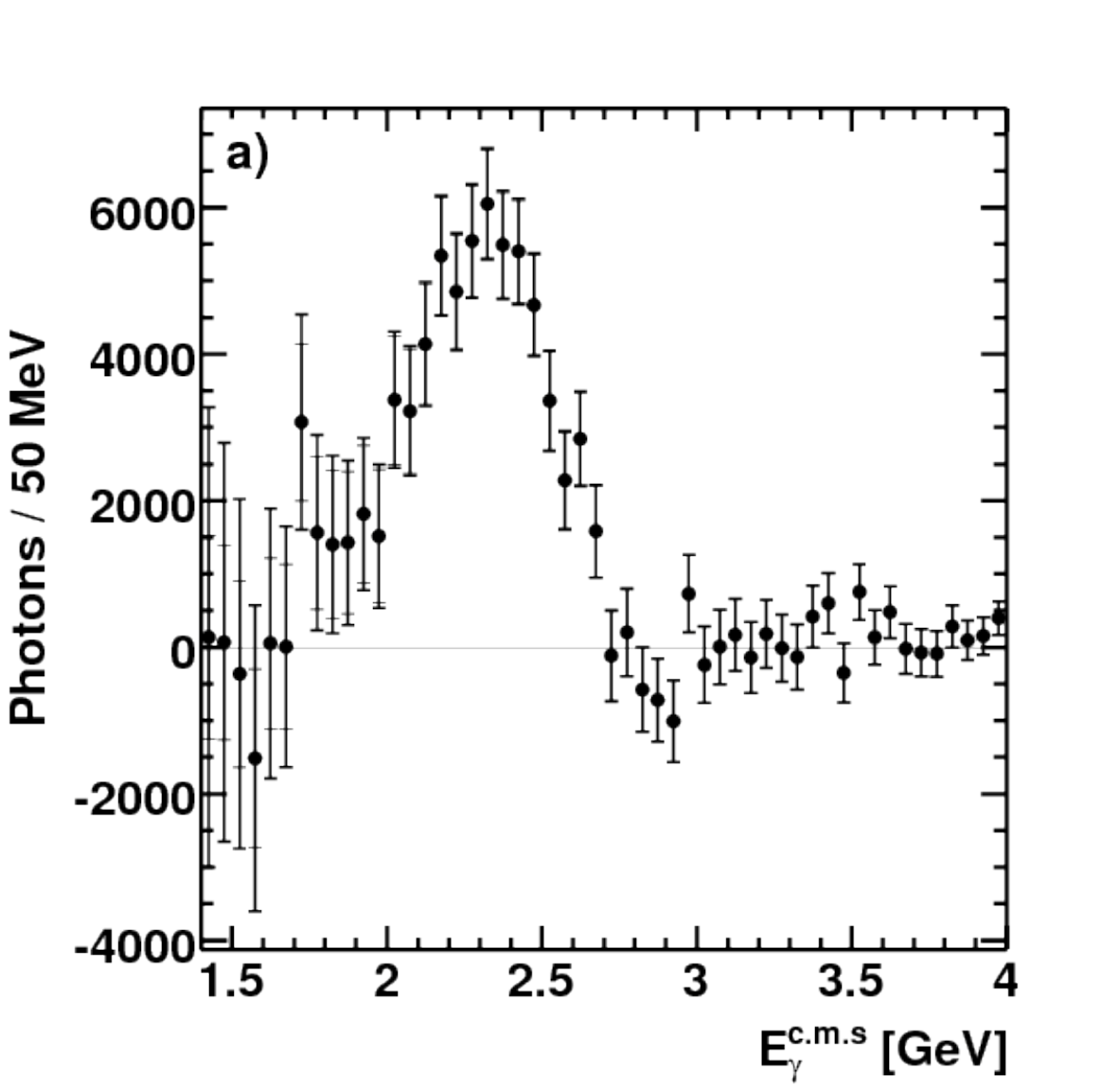}%
\caption{Background-subtracted photon energy spectrum of \bgs decays %
at Belle~\cite{Limosani:2009qg}. \label{Fig:Bgs}}%
  \end{minipage}
\vskip -1.5em ~
\end{wrapfigure}
Only the photon and properties of the rest of the collision event are used to separate 
signal and backgrounds, mostly originating from non-\B decays. At the \B factories 
they are determined from data taken 
at centre-of-mass energies below the $\PUpsilon(4S)$ resonance mass, decays to \piz and $\eta$ which are vetoed and modelled from data, and mis-identified photons which are modelled from simulation.
This method has the disadvantage of larger backgrounds, but has the advantage not to rely on any
modelling of the composition of the hadronic state. It thus comes with smaller theoretical
uncertainties. A typical result for the measured photon energy spectrum in inclusive
\bgs decays is shown in Fig.~\ref{Fig:Bgs}. The integral of the spectrum
gives the decay rate. The width is related to the momentum of the \bquark quark in the \B meson,
which can be seen as a two-body system of a \bquark and a light quark, similar to a
hydrogen atom.

The world-average measured branching fraction for a photon with an energy 
above $1.6\gev$ in the \B-meson rest-frame is 
$(3.32\pm0.15)\times10^{-4}$\mycite{Amhis:2019ckw}. 
It can be compared with
the latest theoretical calculation of $(3.36\pm0.23)\times10^{-4}$\mycite{Misiak:2015xwa,Czakon:2015exa}. 
The very good agreement of these two values sets very strong constraints on New Physics models, 
in particular supersymmetry.
Although the total branching fraction seems to indicate no discrepancy with the Standard Model prediction, 
small contributions from New Physics may still occur. The \bgs decay rate is essentially 
a measurement of the Wilson coefficient $C_7$ (see Section~\ref{Sec:Hamilton}). 
\end{frameenv}


A promising way of searching for right-handed contributions to \bgs decays is the study of time-dependent \CP violation in transitions to flavour eigenstates\mycite{Buras:2015}.
For instance in the decay \decay{\Bz}{\Kstarz\gamma} with \decay{\Kstarz}{\KS\piz}, the amplitude of \decay{\Bz}{\Kstarz\gamma}
may interfere with that of the same decay following \Bz mixing, \decay{\Bd\to\Bdb}{\Kstarz\gamma}.
But this is only the case if the photons have the same helicities, and so the left-handed component of the \Bd decay
will interfere with the right-handed component of the \Bdb decay. This process has been studied at the BaBar and
Belle experiments\mycite{Ushiroda:2006fi,Aubert:2008gy}, and the \Bs counterpart 
decay \decay{\Bs}{\phi\gamma} at LHCb\mycite{LHCb-PAPER-2019-015}. No  significant \CP asymmetry is observed to date.

\subsection[The decays \Bsmm and \Bdmm]{\boldmath The decays \Bsmm and \Bdmm}\label{Sec:Bsmm}
%
\begin{figure}[t]\centering
  \includegraphics[height=0.10\textwidth]{bll_penguin.pdf}
  \hskip 0.03\textwidth
  \includegraphics[height=0.10\textwidth]{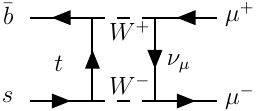}
  \hskip 0.03\textwidth
  \includegraphics[height=0.10\textwidth]{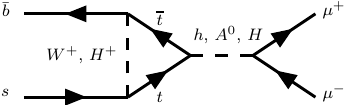} 
  \vskip 0.5em
  \caption{Feynman diagram of the (left and middle) dominating Standard Model contributions to \Bsmm
  and (right) a potential contribution in the context of supersymmetry\mycite{Arbey:2012ax}.}\label{Fig:Bsmm}
\end{figure}
The rare decay \Bsmm proceeds in the Standard Model by a box-type
 diagram involving the \W and \Z bosons and the \tquark quark (Fig.~\ref{Fig:Bsmm}). 
The most recent Standard Model determination of its branching
fraction is $(3.57\pm0.17)\times10^{-9}$\mycite{Beneke:2017vpq,Bobeth:2013uxa,Fleischer:2017ltw},
where the uncertainty is dominated about equally 
by CKM matrix elements and the \Bs decay constant.
\arXiv{In this calculation the branching fraction is evaluated as an average 
over all decay times\mycite{DeBruyn:2012wj,DeBruyn:2012wk}.}{}
In Standard Model extensions, the branching fraction of \Bsmm could be enhanced,
in particular in models containing additional Higgs bosons (Fig.~\ref{Fig:Bsmm}, right). 
The decay \decay{\Bs}{\mumu} and the even more suppressed decay 
\Bdmm have been searched for over three decades, with most recent
results from the 
Tevatron\mycite{Aaltonen:2013as,Abazov:2013wjb} and the 
LHC\mycite{LHCb-PAPER-2011-004,LHCb-PAPER-2011-025,Chatrchyan:2011kr,LHCb-PAPER-2012-007,Chatrchyan:2012rga,ATLAS-CONF-2013-076,LHCb-PAPER-2012-043,LHCb-PAPER-2013-046,Chatrchyan:2013bka,LHCb-PAPER-2014-049,LHCb-PAPER-2017-001,Aaboud:2018mst,Sirunyan:2019xdu} (Fig.~\ref{Fig:BsmmHistory}). 

\begin{figure}[b]\centering
\includegraphics[width=0.55\textwidth]{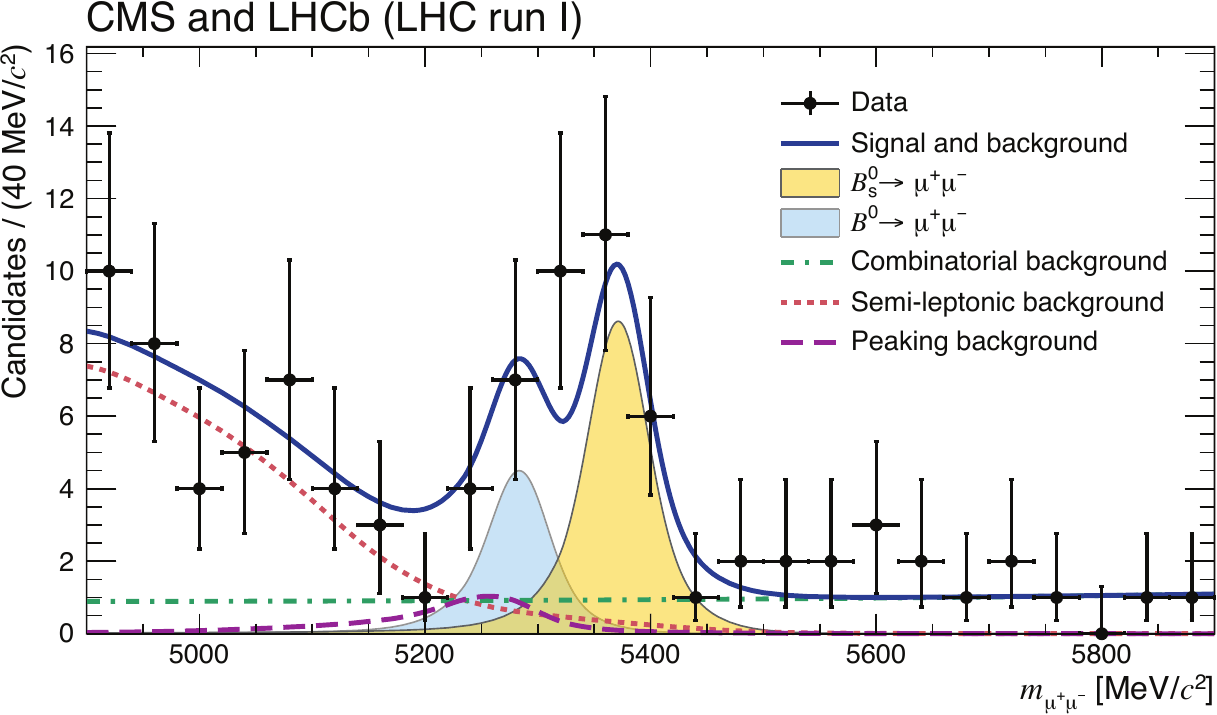}
\includegraphics[width=0.44\textwidth]{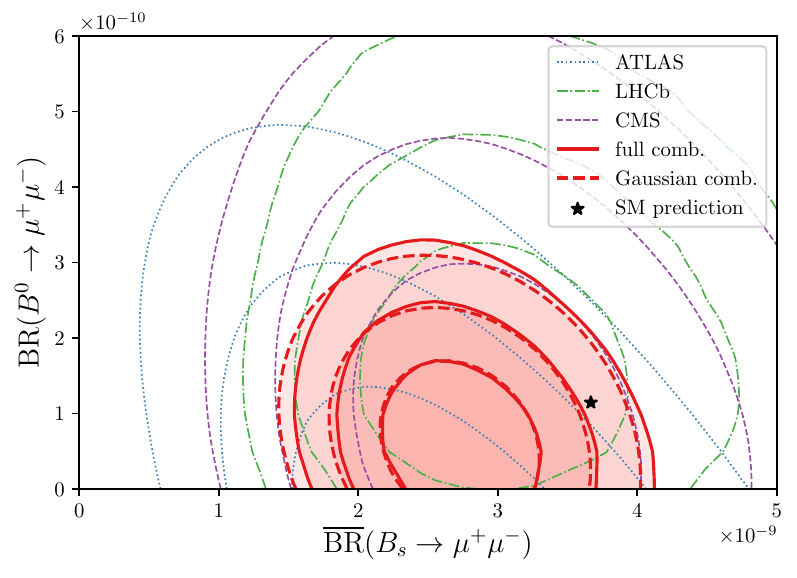}
\caption{(left) Two muon invariant mass distribution of \Bmm candidates 
at CMS and LHCb.
The result of the simultaneous fit is overlaid. Figure from Ref.~\cite{LHCb-PAPER-2014-049}. (right) Experimental constraints on the branching fractions of \Bsmm and \Bdmm.
  Figure, courtesy of D.~Straub, updated from Ref.~\cite{Aebischer:2019mlg}}
\label{Fig:2014-049}\label{Fig:Straub}
\end{figure}

The first observation was reported in 2014 jointly by the CMS and LHCb
collaborations~\cite{LHCb-PAPER-2014-049}. 
The CMS and LHCb LHC Run 1 data sets, which had been already published separately\mycite{LHCb-PAPER-2013-046,Chatrchyan:2013bka}, were combined in a joint fit to 
the data of both experiments. 
The result was published in Nature~\cite{LHCb-PAPER-2014-049},
which is unusual in high energy physics.
The fit to the invariant mass distribution of the two-muon system 
is shown in Fig.~\ref{Fig:2014-049}. The result of the 
combination is an observation of the \Bsmm decay
and a small excess over the background for \Bdmm. 
The first measured branching fraction 
$\BR(\Bsmm)=\left( 2.8 ^{\:+\:0.7}_{\:-\:0.6} \right) \times 10^{-9}$ 
came out consistent with the Standard Model prediction.
For the 
even more suppressed \Bd decay, the result is 
$\BR(\Bdmm)=\left( 3.9 ^{\:+\:1.6}_{\:-\:1.4} \right) \times 10^{-10}$,
which is slightly, but not significantly, larger than the Standard Model prediction
of $(1.06\pm 0.09)\times 10^{-10}$\mycite{Bobeth:2013uxa}.

LHCb, ATLAS and CMS updated their Run~1 result, adding data obtained
in 2015 and 2016\mycite{LHCb-PAPER-2017-001,Aaboud:2018mst,Sirunyan:2019xdu}.
The obtained \Bsmm branching fractions and \Bdmm limits are all consistent.
A graphical representation of all results is provided in Fig.~\ref{Fig:Straub},
with the correlation of the two branching fractions taken into account.

The decays \Bmm are the flagship \Bll modes because of the clean signature provided by the muon pairs. However, \decay{\B}{\epem} and \decay{\B}{\taup\taum} also exist. 
According to its Standard Model prediction, the decay \decay{\Bs}{\epem} is out of reach in the foreseeable future.
Due to the low electron mass, it is even more helicity suppressed than \Bsmm.
Moreover the study of final states involving electrons at hadron colliders is difficult 
due to the lower reconstruction efficiency and the poorer mass resolution 
(see for instance Fig.~\ref{Fig:LHCb-PAPER-2019-009}).
When passing through matter, electrons radiate a significant amount of energy by bremsstrahlung.
This affects the reconstructed momentum and thus smears all derived quantities,
like the invariant mass of the two-electron system.
The best limits are currently $\BF(\decay{\Bs}{\epem}<9.4\times10^{-9}$ and $\BF(\decay{\Bd}{\epem}<2.5\times10^{-9}$ at 90\% C.L.\mycite{LHCb-PAPER-2020-001}. Either limit is obtained assuming the absence of the other mode, as the signals would not be distinguishable due to the poor mass resolution. These limits are fair form the Standard Model, but start to set constraints on models allowing for different couplings to leptons\mycite{Fleischer:2017ltw}, see Sec.~\ref{Sec:LU}.

The expected rate of \decay{\Bs}{\taup\taum} is 
considerably larger, but the decay is experimentally challenging due to the difficult 
\Ptau lepton reconstruction and associated large backgrounds. 
LHCb published a search for the decay \decay{\Bs}{\taup\taum}\mycite{LHCb-PAPER-2017-003},
but with a sensitivity still far from the Standard Model expectation.
Belle II are likely to 
perform improved searches of such decays in the near future.

\subsection{Other leptonic decays}\label{Sec:Blnu}
Just as \Bsmm and \Bdmm are theoretically clean decays, so are
their counterparts with neutrinos. The challenge is on the experimental side.
The decay \decay{\Bz}{\neu\neub} is traditionally labelled as 
``\decay{\Bz}{\text{invisible}}'' as there is no way to experimentally 
determine the number of 
neutrinos (or if there were any at all). In the Standard Model
the branching fraction is vanishing as it is helicity suppressed 
by a factor $(m_\neu/m_\Bz)^3$. Helicity suppression occurs because of
the \Bz meson is spinless, so the two spin-$\sfrac{1}{2}$ neutrinos must have opposite spins.
For massless neutrinos this would be impossible as neutrinos are always left-handed
and antineutrinos right-handed. Only the minute mass of neutrinos
(rarely) allows opposite-spin neutrinos to be emitted. 

Searches have been performed by the \B factory experiments using the 
full reconstruction technique (also referred to as ``on the recoil'').
One \B meson from the \B{}\Bbar pair is fully reconstructed and 
the other is required to leave no trace in the detector. 
The branching fraction is limited to be less than $2.4\times 10^{-5}$ at 90\% confidence level, 
by the \babar experiment\mycite{Lees:2012wv}.

Decays to one charged lepton, \decay{\Bp}{\ell^+\nu}, 
are similarly helicity suppressed, with the 
strength of this suppression depending on the mass of the charged lepton. 
These are tree decays  where the \bquarkbar and \uquark quarks in the \Bp meson annihilate,
but rare  because of the helicity suppression. 
Contributions with the \Wp mediator replaced by a charged Higgs
boson  could enhance or suppress the branching fraction. These decays have all
been searched for by the \B factories using the full reconstruction 
technique describe above.

The \decay{\Bp}{\taup\nu_\tau} decay, where the suppression is the weakest, 
has a predicted branching fraction of
$\BR^\text{SM}=(0.76\aerr{0.08}{0.06})\times 10^{-4}$\mycite{Charles:2015gya} in
the Standard Model and a measured rate of 
$(1.14\pm0.27)\times 10^{-4}$\mycite{PDG,Lees:2012ju,Adachi:2012mm,Hara:2010dk,Aubert:2009wt}, which are in agreement. 
The more suppressed decays
\decay{\Bp}{\ep\nu} and \decay{\Bp}{\mup\nu} have not been observed yet, 
with limits on their branching fraction around 
$10^{-6}$\mycite{Satoyama:2006xn,Aubert:2009ar}.

\subsection[The decays \blls]{\boldmath The decays \blls}\label{Sec:BllKs}\label{Sec:blls}
\begin{figure}[t]%
\includegraphics[width=0.30\textwidth]{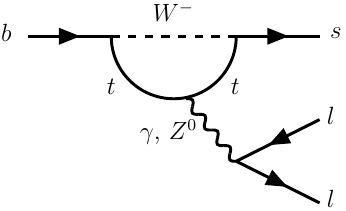}\hskip 0.05\textwidth%
\includegraphics[width=0.30\textwidth]{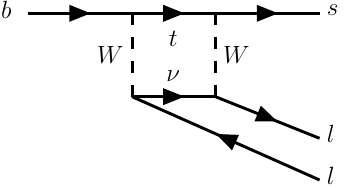}\hskip 0.05\textwidth%
\includegraphics[width=0.30\textwidth]{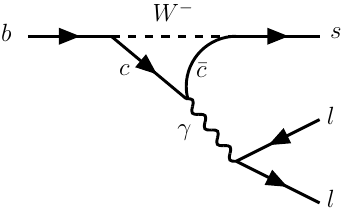}%
\vskip 0.5em
\caption{Feynman diagrams of the dominant Standard Model contributions to \blls: %
(left) electroweak loop, (centre) box, %
(right) \cquark{}\cquarkbar loop diagram.}\label{Fig:blls}%
\end{figure}

The family of decays \decay{\bquark}{\squark\ellp\ellm} ($\ell=e,\mu$) is a laboratory of 
New Physics studies on its own. 
In the Standard Model these decays are induced by a loop diagram similar to that of \bgs 
(but with a \Z component) and a box diagram (Fig.~\ref{Fig:blls}).
The amplitudes corresponding to these diagrams interfere, which causes complex 
phenomenology.

The exclusive decay \decay{\Bz}{\Kstarz\ellp\ellm}, with 
\decay{\Kstarz}{\Kp\pim}, provides a rich set 
of observables with different sensitivities to New Physics, and for which theoretical 
predictions are 
available. These observables are 
affected by varying levels of uncertainties related to the calculation of 
quantum chromodynamical effects. 
Yet, selected ratios of observables benefit from cancellations of 
uncertainties, thus providing a cleaner test of the 
Standard Model\mycite{Eilam:1986fs,Ali:1999mm,Kruger:2005ep,Altmannshofer:2008dz,Egede:2008uy,Bobeth:2008ij,Descotes-Genon:2013vna}. The best known example is the lepton forward-backward 
asymmetry, explained in more details in Inset~\ref{Ins:P5prime}.

\begin{figure}[b]
\begin{minipage}{0.5\textwidth}
\includegraphics[width=\textwidth]{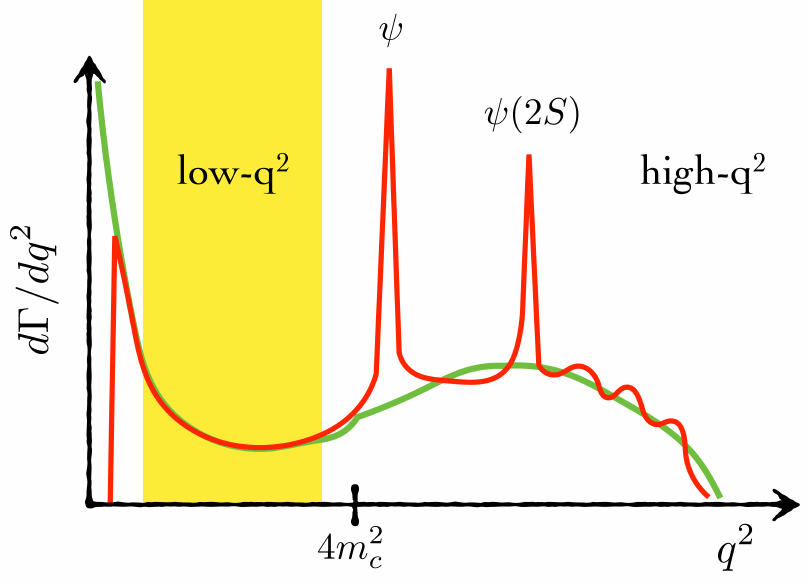}%
\end{minipage}\hskip 0.05\textwidth\begin{minipage}{0.45\textwidth}
\caption{Sketch of the \blls dilepton mass squared distribution, 
courtesy of U. Haisch. 
The yellow band indicates the theoretically favoured 
region $1<q^2<6\:\gevgevcccc$. The green line is the Standard Model contribution from 
operators ${\cal O}_{7,9,10}$ while the red line shows the effect of taking lepton 
masses and \cquark{}\cquarkbar contributions into account.}
\label{Fig:Anatomy}
\end{minipage}
\end{figure}

\begin{frameenv}[t]{\boldmath The $\AFB$, $F_{\text{L}}$ and $P_5'$ asymmetries}\label{Ins:P5prime}
\begin{wrapfigure}{l}{0.4\textwidth}%
\vskip -2em
\includegraphics[width=0.4\textwidth]{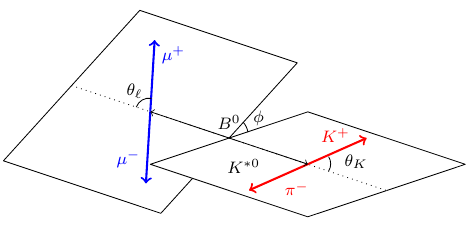}%
\caption{The angles $\theta_\ell$, $\theta_K$ and $\phi$ in the decay \BmmKs. Figure by T.~Blake.}\label{Fig:BllKsAngles}%
\centerline{\includegraphics[width=0.3\textwidth]{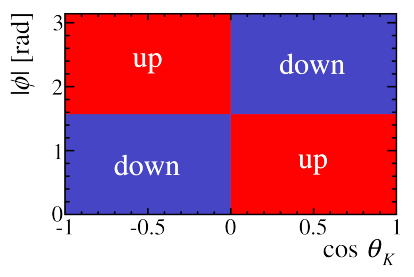}}%
\caption{Definition of the $P_5'$ asymmetry. }\label{Fig:P5prime}
\end{wrapfigure}
In the decay \BdmmKs, followed by \mbox{\decay{\Kstarz}{\Kp\pim}}, 
the direction of the four outgoing particles 
can be described by three angles, shown in Fig.~\ref{Fig:BllKsAngles}. 
The forward-backward asymmetry \AFB is defined as the relative difference between the number of 
positive and negative leptons going along the direction of the \Bz meson in the rest frame 
of the two-lepton system. This corresponds to an asymmetry in the distribution of 
the $\theta_\ell$ angle. 
Similarly, the \Kstarz polarisation fraction $F_\text{L}$ 
depends on the angle $\theta_K$, defined analogously to $\theta_\ell$.

Other asymmetries can be constructed
from the other angles or combinations of them. The $P_5'$ 
asymmetry\arXiv{ suggested by Ref.~\cite{Descotes-Genon:2013vna}}{} 
is based on the angles
$\theta_K$ and $\phi$. It is defined as the relative difference between the number of decays in the 
regions in red and blue in Fig.~\ref{Fig:P5prime}, divided by $\sqrt{F_\text{L}(1-F_\text{L})}$.
Quantities based on several angles are more difficult to measure than single-angle ones 
as they require a better understanding of the reconstruction efficiencies depending on
the kinematics of the outgoing particles.
\end{frameenv}


This interesting picture is complicated by a dependence on $q^2$, 
the dilepton mass squared 
(Fig.~\ref{Fig:Anatomy}).
At very low $q^2$, \BdllKs behaves like \BgKsz, with a slightly off-shell photon 
decaying to two leptons. 
The physics is dominated by the ${\cal O}_7$ operator, as discussed in Section~\ref{Sec:bgs}. 

At higher \qsq values, there is an interference of the amplitudes controlled by the ${\cal O}_{9}$ and ${\cal O}_{10}$ operators,  related to the \Z loop and \W  box diagrams, respectively. 
This ``low-$q^2$'' region between 1 and 6\:\gevgevcccc is the most 
interesting and theoretically cleanest.
Beyond this, non-suppressed \cquark{}\cquarkbar contributions (Fig.~\ref{Fig:blls}, right)
make the picture more 
complicated and theoretical 
estimates are less reliable. 
The observation of high mass resonances above the \psitwos meson
by the LHCb collaboration\mycite[LHCb-PAPER-2013-039]{LHCb-PAPER-2013-039} is an indication that a lot of care is needed when interpreting the high-$q^2$ region.

\begin{figure}[b]
\centering
\includegraphics[width=0.49\textwidth]{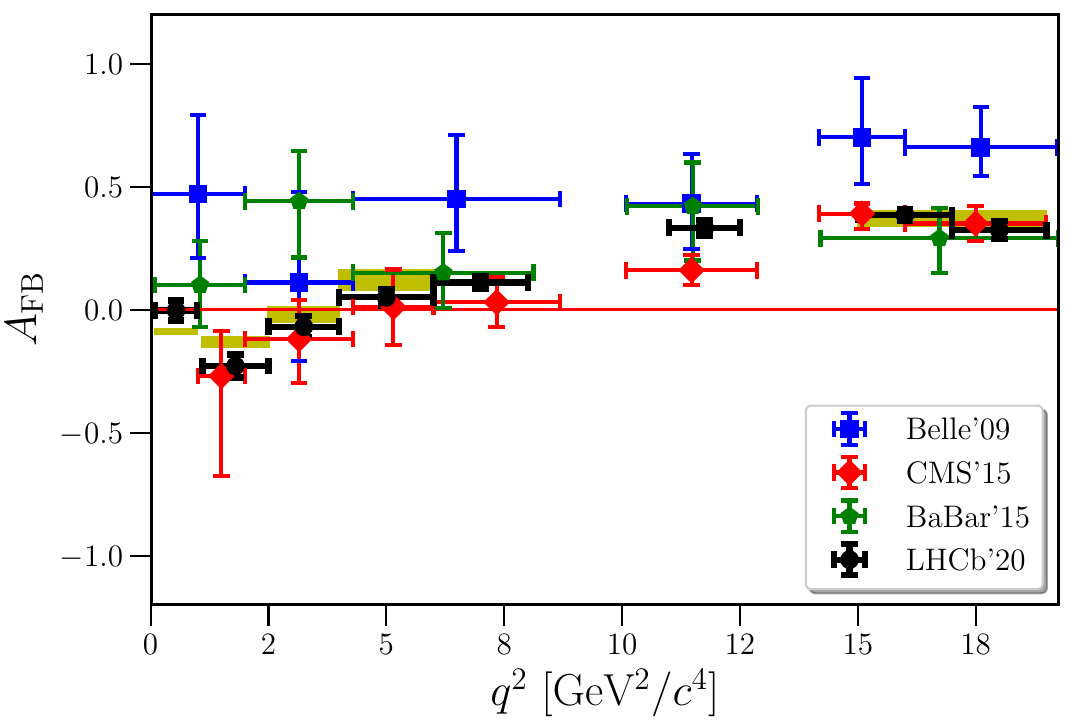}\hskip 0.01\textwidth
\includegraphics[width=0.49\textwidth]{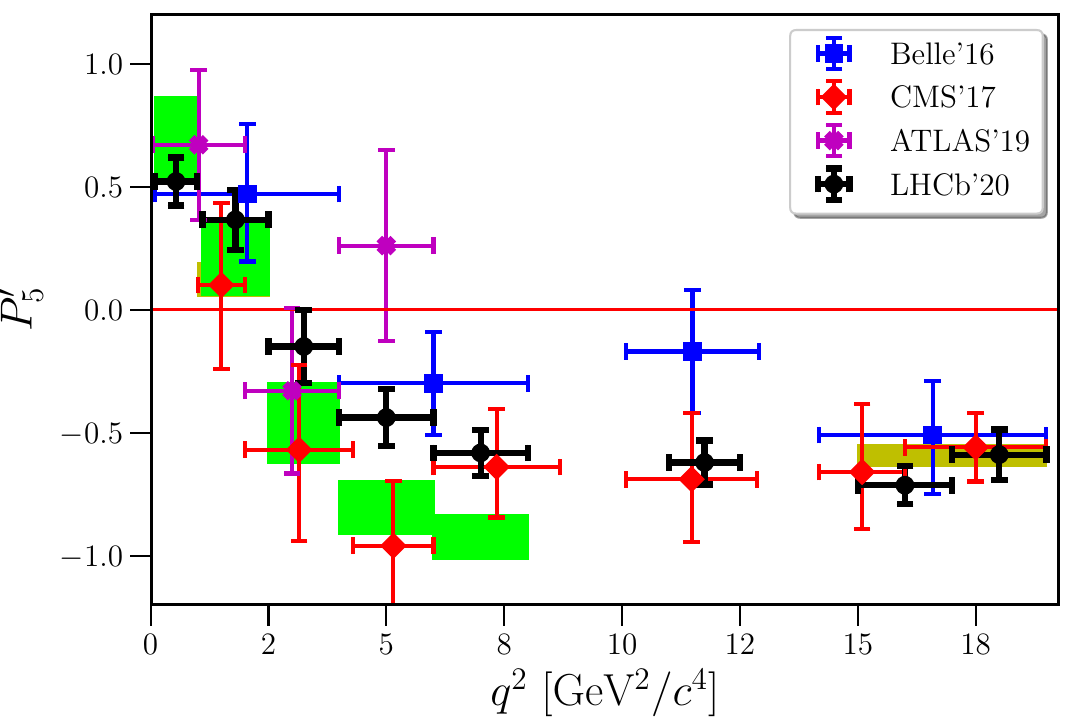}
\arXiv{\caption{Experimental results on the (left) \AFB{}\mycite{Wei:2009zv,Lees:2015ymt,Khachatryan:2015isa,LHCb-PAPER-2020-002} 
and (right) $P_5'$\mycite{Wehle:2016yoi,Sirunyan:2017dhj,Aaboud:2018krd,LHCb-PAPER-2020-002} asymmetries compared to 
theoretical predictions based on the Standard Model\mycite{Straub:2015ica,Altmannshofer:2014rta} (sea green) and\mycite{Descotes-Genon:2014uoa,Khodjamirian:2010vf} (green).}
  }{
\caption{Experimental results for the (left) \AFB{}\mycite{Wei:2009zv,Lees:2015ymt,Khachatryan:2015isa,LHCb-PAPER-2020-002} 
and (right) $P_5'$\mycite{Wehle:2016yoi,Sirunyan:2017dhj,Aaboud:2018krd,LHCb-PAPER-2020-002} asymmetries compared to 
theoretical predictions based on the Standard Model (bands).}
}
\label{Fig:PAPER-2020-002}
\end{figure}

The differential decay width with respect to the dilepton mass squared $q^2$, 
the forward-backward asymmetry $A_\text{FB}$, and the longitudinal 
polarisation fraction $F_\text{L}$ of the \Kstarz resonance have been measured by 
many experiments\mycite{Aubert:2006vb,Aaltonen:2011ja,Aaltonen:2011cn,Chatrchyan:2013cda,ATLAS:2013ola,Khachatryan:2015isa,Lees:2015ymt,LHCb-PAPER-2013-019,Wei:2009zv,LHCb-PAPER-2015-051,LHCb-PAPER-2020-002} with no significant 
sign of deviations
from the Standard Model expectation. 
The most recent measurement of \AFB by the LHCb experiment is shown 
in Fig.~\ref{Fig:PAPER-2020-002} (left). LHCb also studied other angular asymmetries.
In particular in 2013 a local deviation of the $P_5'$ observable (see Inset~\ref{Ins:P5prime})
from the Standard Model expectation was observed around
$q^2\sim5\:\gevgevcccc$\mycite{LHCb-PAPER-2013-037} and then confirmed with larger data sets~\cite{LHCb-PAPER-2020-002}. 
Belle, ATLAS and CMS have subsequently presented data that are consistent with the LHCb 
results\mycite{Wehle:2016yoi,Sirunyan:2017dhj,Aaboud:2018krd}, see Fig.~\ref{Fig:PAPER-2020-002}.


This deviation triggered a lot of interest among theorists\arXiv{, see Refs.~\cite{Gauld:2013qja,Descotes-Genon:2013wba,Hurth:2013ssa,Altmannshofer:2013foa,Datta:2013kja,Mahmoudi:2014mja} for a small subset.}{.}
It is not clear yet if the discrepancy in $P_5'$ 
is a statistical fluctuation, is due to under-estimated 
theoretical uncertainties\mycite{Jager:2012uw,Jager:2014rwa,Beaujean:2013soa,Ciuchini:2015qxb,Hurth:2017hxg,Bobeth:2017vxj}, 
or is the manifestation of a new vector current beyond the Standard Model. 


Similar measurements have been made in the decays 
\mbox{\decay{\B}{\kaon\ellell}}, 
\mbox{\decay{\Bs}{\Pphi\mumu}}, 
\mbox{\decay{\Lb}{\Lz\mumu}} and
\mbox{\decay{\Lb}{\proton\Km\mumu}}\mycite{LHCb-PAPER-2016-059,LHCb-PAPER-2014-032,Wei:2009zv,Aubert:2008ps,LHCb-PAPER-2015-023,Aaltonen:2011cn,LHCb-PAPER-2015-009,Aaltonen:2011qs}. 
The angular observables are consistent with the Standard Model, 
but there is some tension in the branching fraction measurements,
which are on the low side compared to the expectation.

\label{Sec:bnns}
The decay family \bnns is theoretically cleaner than its charged-lepton counterpart \blls.
There are no interferences from \cquark{}\cquarkbar loops as those
do not annihilate to neutrino pairs. The main difficulty 
is on the experimental side and only \B factory experiments have attempted
looking at such decays using the full reconstruction technique.
None have been found and the most stringent limits on the decay rates 
of \decay{\Bz}{\Kstarz\neu\neub} and \decay{\Bp}{\Kp\neu\neub} 
are at the $10^{-5}$ level\mycite{Lees:2013kla}.

\subsection{Lepton universality tests}\label{Sec:LU}
Most of the above-mentioned \blls 
measurements assume that muons and electrons behave the same way.
This assumption, called {\it lepton universality}, is built into the Standard Model
and has been extensively tested, most notably at LEP experiments. The
only  Standard Model particle that has different couplings to leptons is the Higgs boson, which 
couples proportionally to mass. The presence of new particles that would
couple differently to leptons can be tested by measuring the ratio
\begin{equation}
  R_H \equiv \frac{{\cal B}(\decay{\B}{H\mumu})_\qsq}{{\cal B}(\decay{\B}{H\epem})_\qsq},
\end{equation}
where $H$ is any hadron and the \qsq index indicates that this ratio is to be measured in a well-suited range of dilepton masses.

\begin{figure}[t]
\centering
\includegraphics[width=0.49\textwidth]{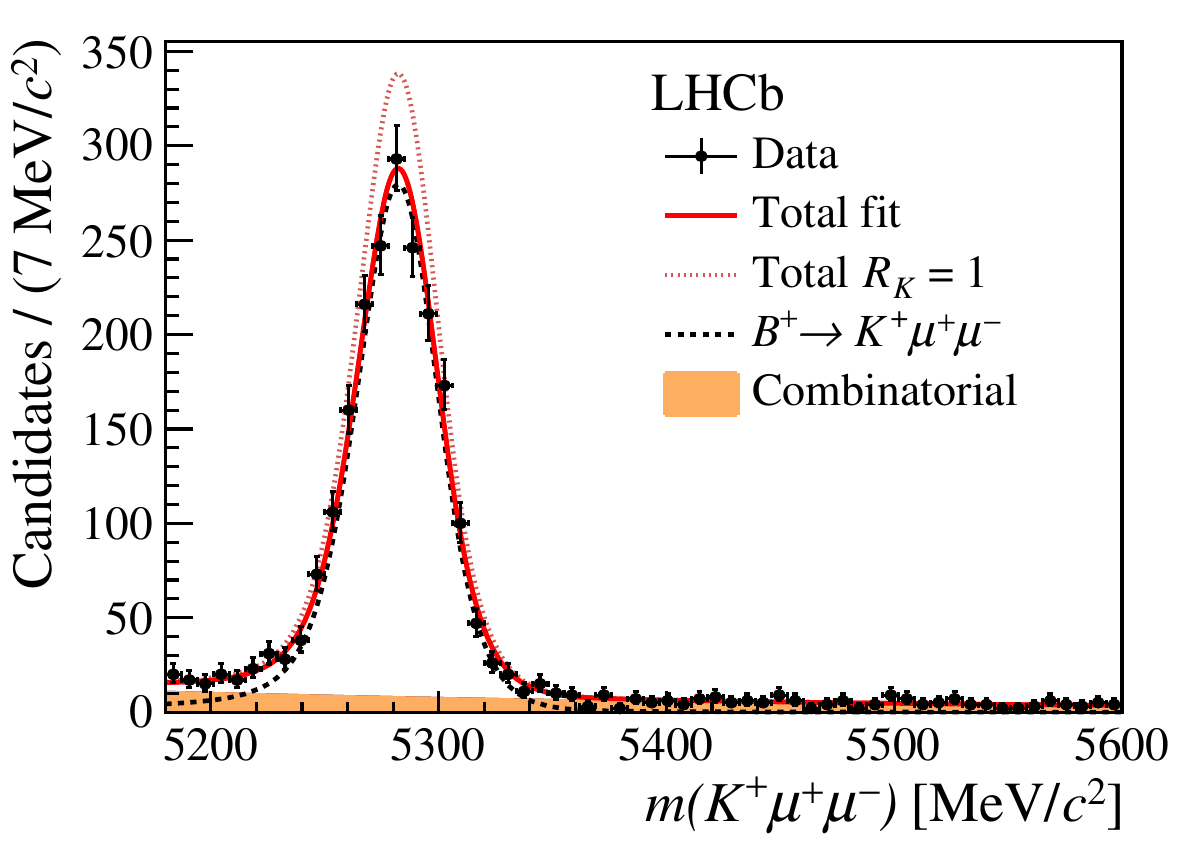}\hskip 0.01\textwidth
\includegraphics[width=0.49\textwidth]{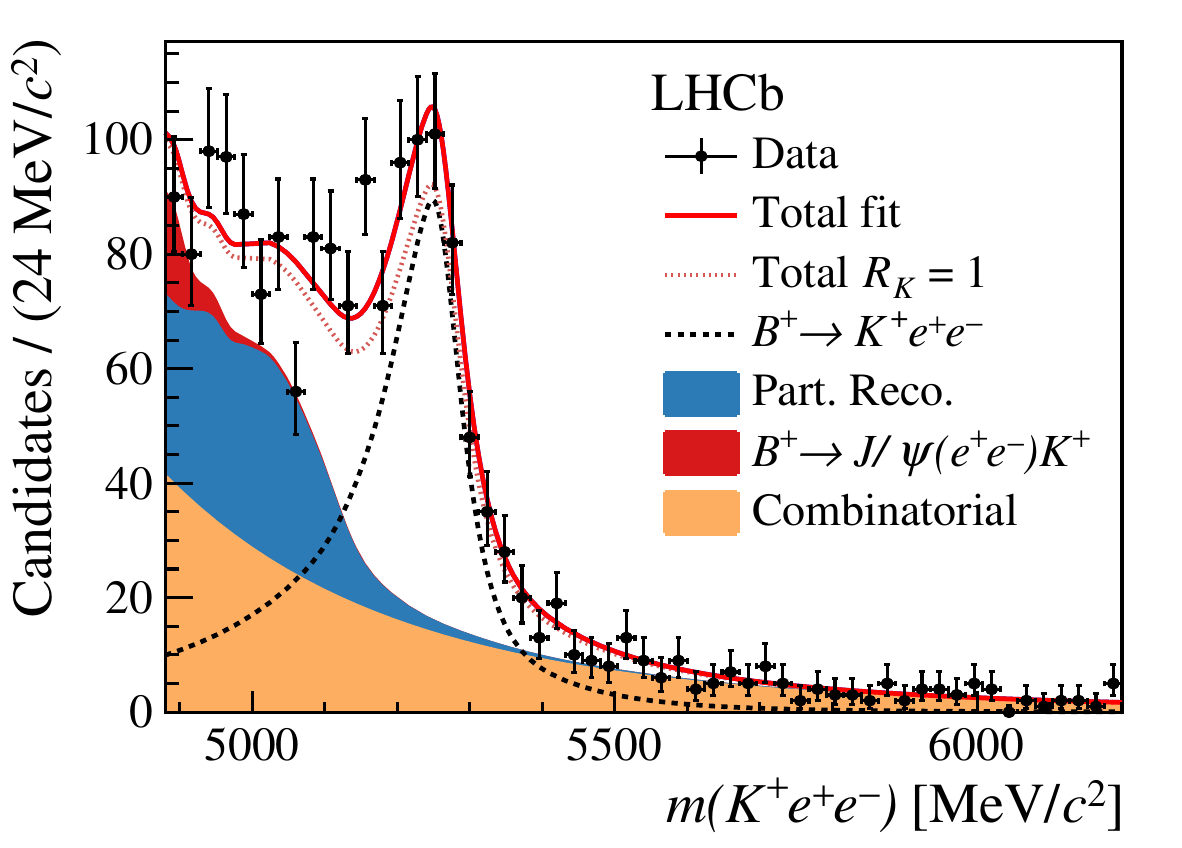}\\
\caption{Mass distributions of (left) \BummK and (right) \BueeK.
Figures from Ref.~\cite{LHCb-PAPER-2019-009}. The light-shaded are shows the
combinatorial background, and the dark-shaded the background from
partially reconstructed \decay{\B}{\Kp\ellell X} decays where $X$ is undetected.}
\label{Fig:LHCb-PAPER-2019-009}
\end{figure}
Surprisingly, the lepton universality ratio in \BllKp decays,
$R_K$,
was measured to be lower than unity with LHCb Run 1 data\mycite{LHCb-PAPER-2014-024}
in the $1<q^2<6\:\gevcc$ range, which is most sensitive to New Physics contributions.
The Standard Model prediction for this ratio is unity within 
$10^{-3}$\mycite{Hiller:2003js,Bobeth:2007dw}
as all hadronic uncertainties cancel in the ratio.
The experimental value was then updated with 2016 data to 
$\RK=0.846\aerr{0.060}{0.054}\aerr{0.016}{0.014}$\mycite{LHCb-PAPER-2019-009},
which corresponds to a $2.5\sigma$ tension with unity. 
Fig.~\ref{Fig:LHCb-PAPER-2019-009} shows the mass
peaks of \BmmKp and \BeeKp, highlighting the effect
of bremsstrahlung affecting electron reconstruction. 
A similar deviation is seen by \babar{}\mycite{Lees:2012tva} and
\belle{}\mycite{Abdesselam:2019lab} although with 
larger uncertainties.

\begin{figure}[t]
\centering
\includegraphics[width=0.49\textwidth]{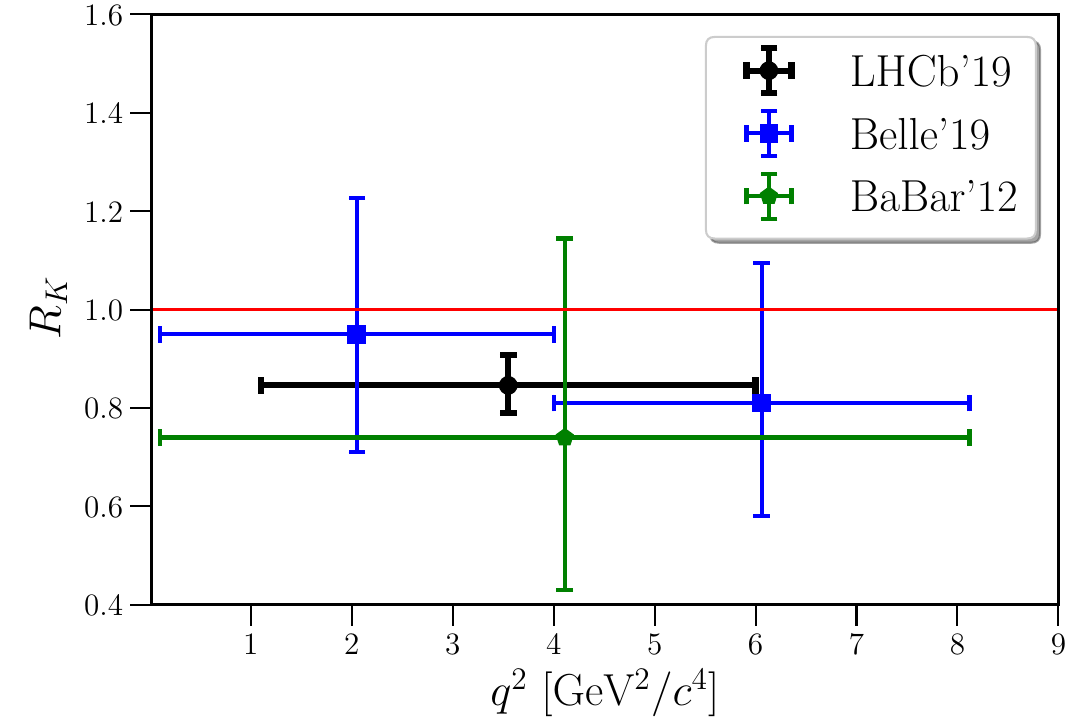}\hskip 0.01\textwidth
\includegraphics[width=0.49\textwidth]{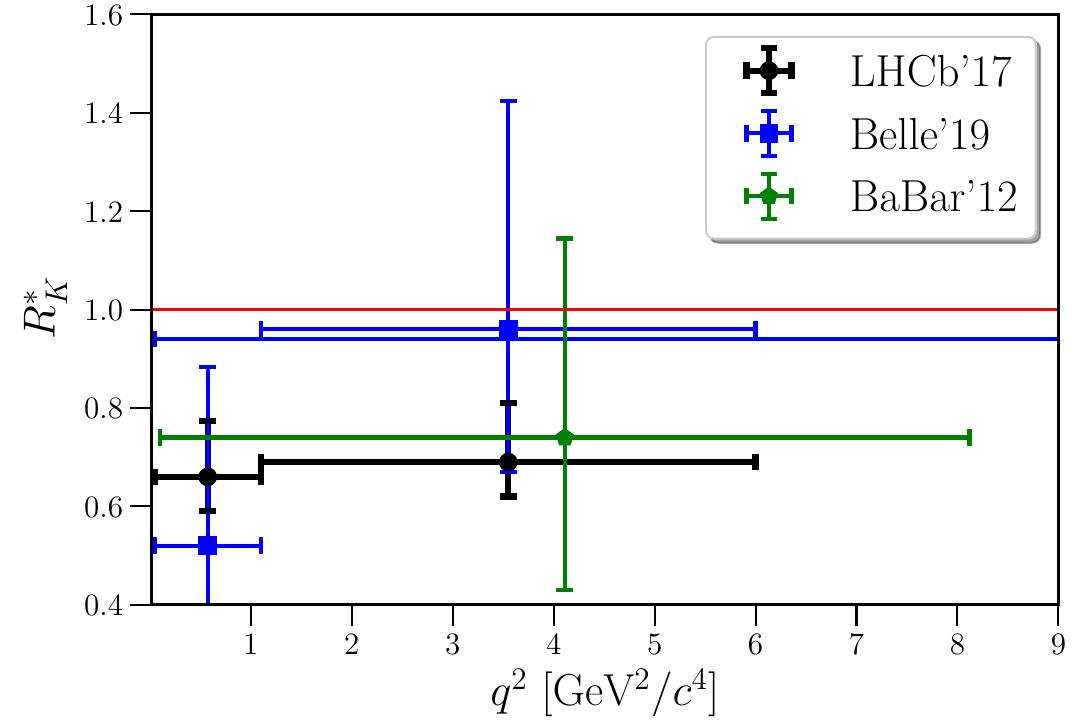}\\
\caption{Measurements of (left) \RK and (right) \RKstar in the low-\qsq region
  from LHCb\mycite{LHCb-PAPER-2019-009,LHCb-PAPER-2017-013}, BaBar\mycite{Lees:2012tva} and Belle\mycite{Abdesselam:2019lab,Abdesselam:2019wac}.}
\label{Fig:RK}
\end{figure}

The ratio $R_\Kstar$ is defined in analogy. LHCb published two
measurements in bins of $q^2$\mycite{LHCb-PAPER-2017-013}, which each differ
by about $2\sigma$ from the Standard Model expectation of
approximately unity\mycite{Jager:2014rwa,Altmannshofer:2017fio,Serra:2016ivr,Capdevila:2017ert,Bordone:2016gaq}.
These results are more precise than those determined by the
Belle\mycite{Abdesselam:2019wac} and BaBar\mycite{Lees:2012tva} collaborations,
as shown in Fig.~\ref{Fig:RK}.

Similar decays can be used to perform tests of lepton universality in 
\mbox{\decay{\Bs}{\phi\ellell}},
\mbox{\decay{\Lb}{\Lz\ellell}},
\mbox{\decay{\Lb}{\proton\Km\ellell}}, 
which are all accessible by the LHCb experiment, but some
have very limited yields. 
These measurements
are complementary, as the different
spins of the hadronic component probe different New Physics couplings\mycite{Hiller:2014ula}.
Also, the angular distributions described in Sec.~\ref{Sec:blls} should be 
investigated separately for decays to electrons and to muons. 
Belle reported separately the values of the $P_5'$ asymmetry\mycite{Wehle:2016yoi},
but no discrepancies
were observed given the small available data sample.

\label{Sec:tau}
Given the hints of lepton-flavour universality violation between muons and electrons,
it seems natural to wonder if such an effect
can be seen in processes involving the third generation (\Ptau) lepton.
This has been tested in \decay{\B}{D^{(\ast)}\tau\neub_\tau} decays, 
comparing to the same decay with muons or electrons instead of tau leptons.
Unlike the decays described above, 
the Standard Model contribution to this decay is not suppressed (and does not match the definition 
of a rare decay). It proceeds via a tree-level \decay{\bquark}{\cquark\Wm} transition, with the 
\Wm decaying to a lepton and a neutrino. The expectation is that the rates 
for the decays involving electrons, muons and tau leptons differ only due to 
phase-space effects (plus small effects due to form factors).
The ratios of the rates of \decay{\bquark}{\cquark\tau\neub} to \decay{\bquark}{\cquark\ell\neub} 
($\ell=\mu,e$) measured by BaBar\mycite{Lees:2012xj,Lees:2013uzd}, 
Belle\mycite{Huschle:2015rga,Hirose:2016wfn,Hirose:2017dxl,Abdesselam:2019dgh} and LHCb\mycite{LHCb-PAPER-2015-025,LHCb-PAPER-2017-017,LHCb-PAPER-2017-027}
come out larger than expected, with an average~\cite{Amhis:2019ckw} deviating by 
approximately $3\sigma$ from the 
Standard Model predictions\mycite{Aoki:2016frl,Na:2015kha,Bailey:2012jg,Fajfer:2012vx}.
This could indicate the presence of new couplings preferring tau leptons.

\section{Wilson coefficient fits}\label{Sec:Fits}
This section briefly describes some constraints on Wilson coefficients\arXiv{, as of summer 2019}{}.
It relies on Section~\ref{Sec:Hamilton}. 
Several groups have performed model-independent fits of Wilson coefficients, 
using most of the experimental results presented above. 
\arXiv{See Refs.~\cite{Aebischer:2019mlg,Alok:2019ufo,Ciuchini:2019usw,Alguero:2019ptt,Arbey:2019duh,Shi:2019gxi,Capdevila:2019tsi}
for a representative subset.}{} The fits differ by the set of experimental results used, the 
statistical treatment of uncertainties and choices of form factors. Another major difference is the 
level of trust of computations of quark loops (most notably \cquark{}\cquarkbar loops)
incorporated in the fit. Depending on these choices, the determined tension with the Standard 
Model ranges from one to several standard deviations. 

\begin{wrapfigure}{r}{0.46\textwidth}\centering
\begin{minipage}{0.44\textwidth}
\includegraphics[width=\textwidth]{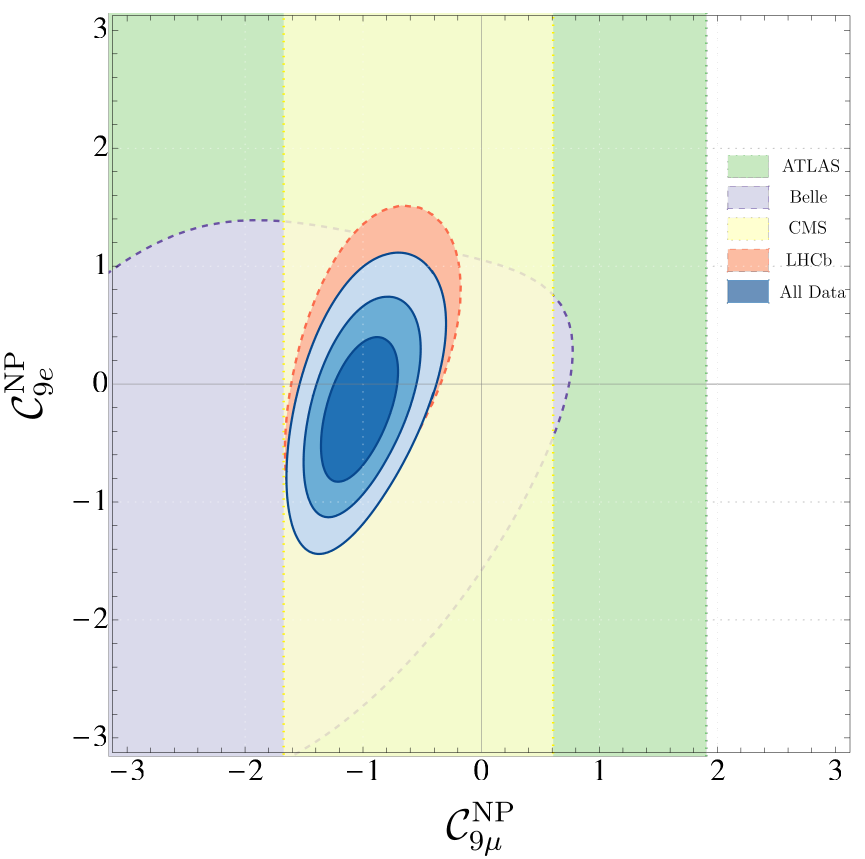}
\caption{Allowed regions in the $C_{9e}^\text{NP}$ versus $C_{9\mu}^\text{NP}$ plane. 
Figure from Ref.~\cite{Alguero:2019ptt}.}\label{Fig:Fits}
\end{minipage}
\vskip -3em
\end{wrapfigure}

In all cases, the New Physics scenario which is preferred changes the value of the $C_9$ coefficient 
(adding a non-zero term $C_9^\text{NP}$). This term could then be different depending on the flavour 
of the involved leptons (introducing $C_{9e}^\text{NP}$ and $C_{9\mu}^\text{NP}$), thus breaking lepton universality, 
see Fig.~\ref{Fig:Fits}.
The data are not conclusive yet, but a tension with the Standard Model point at $(0,0)$ is visible.
The significance of this tension depends on the assumed theory uncertainties.

Another popular model is to assume that the weak interaction $V-A$ structure holds in 
New Physics and thus to impose $C_9^\text{NP}=-C_{10}^\text{NP}$. 
The data are consistent with such
a hypothesis, but again it is too early to draw conclusions.

Right-handed components are also added in the fits, in particular using asymmetries in 
\mbox{\blls} decays that are sensitive to such effects. Presently there is no evidence for any
significant need for right-handed currents.

There is a plethora of model-dependent interpretations of these findings.
The deviations can be 
accommodated by
supersymmetry\mycite{Mahmoudi:2014mja},
models with new vector bosons\mycite{Gauld:2013qja,
  Glashow:2014iga,
  Alonso:2014csa,
  Ghosh:2014awa,
  Altmannshofer:2014cfa,
  Crivellin:2015lwa,
  Celis:2015ara,
  Belanger:2015nma,
  Falkowski:2015zwa,
  Chiang:2016qov,
  Crivellin:2016ejn,
  Hiller:2017bzc,
  Kamenik:2017tnu,
  Ko:2017lzd,
  Sala:2017ihs,
  Kowalska:2019ley}, 
two Higgs doublets\mycite{Crivellin:2015mga,
  Arbey:2017gmh},  
scalar interactions\mycite{Datta:2013kja,
  Arnan:2016cpy,
  Arnan:2019uhr} or 
leptoquarks\mycite{Dorsner:2016wpm,
  Hiller:2014yaa,
  Gripaios:2014tna,
  Fajfer:2015ycq,
  Alonso:2015sja,
  Bauer:2015knc,
  Deppisch:2016qqd,
  Dorsner:2017ufx,
  Becirevic:2016yqi,
  Bordone:2017bld,
  Becirevic:2017jtw,
  Hiller:2017bzc,
  Crivellin:2017dsk,
  Crivellin:2017zlb,
  DiLuzio:2017fdq,
  Buttazzo:2017ixm,
  Becirevic:2018afm,
  Angelescu:2018tyl,
  Kowalska:2019ley}.

\section{Prospects}\label{Sec:Prospects}\label{Sec:Outlook}
At the risk of stating the obvious,
rare decays have the advantage of being rare. This ensures that the experimental precision will stay 
dominated by statistical uncertainties, and thus will not run into a limit imposed by irreducible 
systematic uncertainties. The theoretically cleanest measurements, like the 
lepton-universality ratios $R_{X_s}$ and 
the ratio of \Bdmm to \Bsmm branching fractions will continue to be of interest as more data are 
acquired at the LHC and by Belle II. The future will tell us if the deviations from 
expectations
hold and tell us something new about Nature. 

Other measurements, like branching fractions (for instance \bgs) have already reached the 
theoretical 
precision and more work is needed on this side to allow more precise comparisons of experimental 
values and Standard Model predictions. Finally, asymmetries in \BllKs are in between. If the presently 
measured central values stay while the uncertainties reduce, we may soon be in the situation of 
having to understand a very significant deviation with predictions based on the Standard Model. 
More investigations of theory uncertainties are needed before  any conclusion can be reached.

\arXiv{\section{Conclusion}\label{Sec:Conclusion} 
Rare decays provide a useful tool to search for physics beyond the Standard Model. Many intriguing 
results hinting at New Physics stem from from rare decay measurements at the \B factories or the LHC. 
These measurements do not tells us straight away which kind of New Physics could cause the seen 
deviations, but allow for model-independent analyses describing the common features of possible 
explanations. This is in turn needed for model building.
Recent results, especially about \Bmm and \blls decays, have triggered a lot of new 
models that may be confirmed by the observation of on-shell new particles 
if those are within reach of present colliders.}{}

The analysis of LHC Run 2 data 
is now ongoing and the Belle II experiment has just started. Improved
measurements of the processes described in this article, 
and also new complementary measurements, will become available and 
will lead to improved precision which will be useful in global fits.

\arXiv{\section*{Acknowledgements}
The authors would like to thank 
Lydia Roos, Andreas Hoecker and Tim Gershon
for carefully reading the manuscript
and Adrian Bevan for his useful comments.
}{}

\addcontentsline{toc}{section}{References}
\setboolean{inbibliography}{true}
\bibliographystyle{LHCb}
\bibliography{mystandard,LHCb-PAPER,LHCb-CONF,LHCb-DP,exp,theory,other}

\end{document}